\documentclass[fleqn,usenatbib]{mnras}
\usepackage{savesym}
\usepackage{amsmath}
\savesymbol{iint}
\usepackage[utf8]{inputenc}
\usepackage{graphicx}
\usepackage{txfonts}
\restoresymbol{TXF}{iint}
\usepackage{footnote}
\usepackage{tablefootnote}
\usepackage[para,online,flushleft]{threeparttable}
\usepackage{xcolor}
\usepackage{xfrac}
\usepackage{caption,subcaption}     %subfigures
\definecolor{redak}{rgb}{0.9,0.15,0.05}

\def\kms{\, \mathrm{km}\, {\mathrm s}^{-1}}
\def\cms{\, \mathrm{cm}\, {\mathrm s}^{-2}}
\def\smy{\,\mathrm{M}_\odot$\,${\rm yr}^{-1}}

\defcitealias{deJager1988}{dJ88}
\defcitealias{Vink2001}{V01}
\defcitealias{Vink2017}{V17}

\title[$\upsilon$\,Sgr did it again]{Ups!... I did it again: unveiling the hidden companion in Upsilon Sagittarii, a unique binary system at a second mass transfer stage\thanks{Based on observations made with the Mercator Telescope, operated on the island of La Palma by the Flemish Community, at the Spanish Observatorio del Roque de los Muchachos of the Instituto de Astrofísica de Canarias, and on observations made with the FEROS spectrograph attached to the 2.2-m MPG/ESO telescope at the La Silla observatory.} }
\author[Gilkis \& Shenar]{Avishai~Gilkis\thanks{\href{agilkis@tauex.tau.ac.il}{agilkis@tauex.tau.ac.il}} and Tomer~Shenar\thanks{\href{t.shenar@uva.nl}{t.shenar@uva.nl}}
\\
$^{1}$ The School of Physics and Astronomy, Tel Aviv University, Tel Aviv 6997801, Israel\\
$^{2}$ Anton Pannekoek Institute for Astronomy, Science Park 904, 1098 XH, Amsterdam, The Netherlands \\
}

\pubyear{2022}

\begin{document}
\label{firstpage}
\pagerange{\pageref{firstpage}--\pageref{lastpage}}

\maketitle

% ==========================================================
\begin{abstract}\\

Upsilon Sagittarii is a hydrogen-deficient binary that has been suggested to be in its second stage of mass transfer, after the primary has expanded to become a helium supergiant following core helium exhaustion. A tentative identification of the faint companion in the ultraviolet led to mass estimates of both components that made the helium star in  Upsilon Sagittarii a prototypical immediate progenitor of a type Ib/c supernova. However, no consistent model for the complex spectrum has been achieved, casting doubt on this interpretation. In the present study we provide for the first time a composite spectral model that fits the ultraviolet data, and clearly identifies the companion as a rapidly rotating, slowly moving $\approx 7\,\mathrm{M}_\odot$ B-type star, unlike previously suggested. The stripped helium supergiant is less luminous than previous estimates, and with an estimated mass of $ < 1\,\mathrm{M}_\odot$ is ruled out as a core-collapse supernova progenitor. We provide a detailed binary evolution scenario that explains the temperature and luminosity of the two components as well as the very low gravity ($\log g \approx 1$) and extreme hydrogen deficiency of the primary (atmospheric mass fraction $X_\mathrm{H,1} \approx 0.001$). The best-fitting model is an intermediate-mass primary ($M_\mathrm{ZAMS,1} \approx 5\,\mathrm{M}_\odot$) with an initial orbital period of a few days, and a secondary that appears to have gained a significant amount of mass despite its high rotation. We conclude that Upsilon Sagittarii is a key system for testing binary evolution processes, especially envelope stripping and mass accretion.

\end{abstract}

\begin{keywords}
binaries:~close --- binaries:~spectroscopic --- stars:~evolution -- stars:~individual:$~\upsilon$\,Sagittarii
\end{keywords}
% ==========================================================

% ==========================================================
\section{Introduction}
\label{sec:intro}
% ==========================================================

Hydrogen-deficient supernovae (SNe Ib/c) are thought to be the end products of massive helium stars: pre-collapse stars that were stripped of their outer layers \citep*{Podsiadlowski1992, Smartt2009, Yoon2010}. Massive helium stars have so far been predominantly observed as classical Wolf-Rayet (WR) stars, which are hot, hydrogen-depleted stars that possess powerful stellar winds. However, WR stars are thought to represent the upper-mass end of the predicted helium-star population -- those helium stars with initial masses of at least $M_{\rm i} \gtrsim 20\,\mathrm{M}_\odot$ in our Galaxy \citep{Massey1981, Hamann2019, Shenar2020WR}. While the final fate of such stars is still debated, both theoretical \citep[e.g.,][]{Sukhbold2016} as well as empirical \citep[e.g.,][]{Mirabel2003, Shenar2022} evidence suggest that such stars tend to directly collapse into black holes, without an associated supernova explosion (though see also \citealt{Gal-Yam2022, Mahy2022}). Therefore, massive helium stars of lower masses (i.e., $M_{\rm i} \lesssim 20\,\mathrm{M}_\odot$) are favoured as progenitors of SNe Ib/c. Such objects are not expected to appear as WR stars, and are thought to form via mass transfer in close binaries \citep*{Paczynski1967, Dionne2006, Goetberg2017}. 

From the frequency of binary stripping \citep{Sana2012, Sana2013, deMink2014} and simple lifetime and initial mass function arguments, one can estimate that there are a factor $5$--$10$ more of these lower mass helium stars compared with classical WR stars, amounting to tens of thousands in our Galaxy alone \citep[see also][]{Goetberg2017}. However, only two such helium stars were reported to exist in our Galaxy as of yet: the primaries of the binaries $\upsilon$\,Sgr (HD~181615) and HD~45166 \citep*{Dudley1990, Groh2008}.  The helium star in HD~45166, termed ``quasi Wolf-Rayet'' (qWR), was recently found to be strongly magnetic, and likely did not follow the standard evolutionary scenario responsible for the production of the bulk of SNe Ib/c progenitors (Shenar et al., submitted). This leaves the primary of $\upsilon$\,Sgr, classified A2~I, as the only prototypical SN Ib/c progenitor proposed in the literature.

Unusual spectral features of $\upsilon$\,Sgr were noted more than a century ago. \cite{Campbell1895} included $\upsilon$\,Sgr in the list of ``stars whose spectra contain both bright and dark hydrogen lines''. \cite{Cannon1912} described in detail the peculiarities in its spectrum, and suggested that the system is composed of multiple components. Radial velocity measurements by \cite{Campbell1899} indicated that the system is a binary, and an orbital period of $P\approx 138\,\mathrm{d}$ was derived by \cite{Wilson1915}, essentially identical to modern estimates \citep{Koubsky2006}. The spectrum of $\upsilon$\,Sgr exhibits an extreme hydrogen deficiency in the bright component \citep{Greenstein1940,Hack1966,Hack1980}, and an ultraviolet (UV) excess of a hotter so-called ``invisible'' companion \citep*{Duvignau1979,Parthasarathy1986}. The most recent spectroscopic analysis of $\upsilon\,$Sgr was performed by \cite{Dudley1992}, who however assumed local thermodynamic equilibrium (LTE). This assumption does not necessarily hold in the conditions prevailing in the atmosphere of the primary.

\cite{Schoenberner1983} provided an evolutionary scenario to explain all the observed features of $\upsilon$\,Sgr. First, mass transfer by Roche-lobe overflow (RLOF) occurs after the primary evolves off the main sequence (MS). Most of the hydrogen-rich envelope is lost in this stage, but the primary contracts and ceases RLOF while retaining a low-mass but hydrogen-rich envelope. The primary proceeds to evolve as a ``stripped helium star'', near the helium MS, and expands again after helium is depleted in its core. A second stage of mass transfer by RLOF then reduces the surface hydrogen mass fraction to the observed deficiency. According to \cite{Schoenberner1983}, $\upsilon$\,Sgr is now observed in this mass-transfer stage, with its high luminosity provided by a helium-burning shell surrounding an inert carbon-oxygen core\footnote{A similar scenario was discussed by \cite*{Halabi2018} for systems that have gone through common envelope evolution, but this is not thought to be the case for $\upsilon$\,Sgr.}. This type of hydrogen-deficient binary system in a second mass transfer stage is observationally rare -- two other known similar systems are KS~Persei \citep*[HD~30353;][]{Drilling1982,Parthasarathy1990} and LSS~4300 \citep*[HD~320156;][]{Schoenberner1984}.

\cite{Dudley1990} argued to have inferred radial velocities (RVs) for the faint component of $\upsilon$\,Sgr from spectra obtained with the \textit{International Ultraviolet Explorer} (IUE). According to the mass ratio derived by \cite{Dudley1990}, the minimum mass of the stripped helium-star primary is $M_\mathrm{p,min}=2.52\pm 0.05\,\mathrm{M}_\odot$, indicating that it is a strong candidate progenitor for a Type Ib core-collapse supernova. To achieve this, \cite{Dudley1990} cross-correlated the co-added IUE spectrum with the individual IUE spectra in the range 1230--1380\,\AA. An extraction of the spectrum of the secondary in this spectral range by \citet{Dudley1990} yielded a spectrum that does not appear stellar, containing only few sparse absorption features that are as narrow as a resolution element ($\Delta \varv \approx 10\,\kms$), implying that the companion is a very slow rotator. The unusual, non-stellar appearance of the extracted spectrum of the secondary suggests that the results are spurious rather than real.

\cite{Netolicky2009} used mid-infrared interferometric observations to measure the properties of a thin circumbinary disc, which is assumed to have formed from past binary interaction in the system. The inclination of the disc, $\approx 50^\circ$, is therefore expected to coincide with that of the binary orbit. Combined with the analysis by \cite{Koubsky2006}, the mass of the stripped star becomes even higher than the minimum given by \cite{Dudley1990}. \cite{Koubsky2006} and \cite{Netolicky2009} emphasise that verification of the radial velocity curve of the secondary is needed.

In the present study, we perform a multi-wavelength spectroscopic analysis using atmosphere models and high-resolution spectra of $\upsilon$\,Sgr without assuming LTE (non-LTE). We show that the companion, which we argue has not been identified thus far, is a rapidly rotating star that is 5--10 times more massive than the primary helium star. We find a detailed binary evolution scenario that reproduces the observed properties, and rule out the possibility that $\upsilon$\,Sgr is an immediate core-collapse supernova progenitor.

In Sect.\,\ref{sec:specan} we describe the observational data and spectroscopic analysis. In Sect.\,\ref{sec:StellarEvolution} we describe the main aspects of the stellar evolution simulations and our best-fitting model. In Sect.\,\ref{sec:discussion} we discuss our main findings and in Sect.\,\ref{sec:summary} we summarise.

% ==========================================================
\section{Spectroscopic analysis}
\label{sec:specan}
% ==========================================================

% ==========================================================
\subsection{Observational data}
\label{subsec:obsdat}
% ==========================================================

For the spectroscopic analysis, we use spectra acquired in the years 2011--2013 with the HERMES spectrograph mounted on the 1.2\,m Mercator telescope at the Observatorio del Roque de Los Muchachos on La Palma, Spain \citep{Raskin2011}. HERMES spectra cover the wavelength range from 3770 to 9000\,\AA\ with a spectral resolving power of $R\approx$\,85000. A total of 23 HERMES spectra are available, having a typical signal-to-noise (S/N) of 100 at the continuum. Standard calibrations including bias and flat-field corrections and wavelength calibrations were performed using their respective pipelines. Barycentric correction was applied. We normalised the spectra by dividing the spectra through a piecewise-linear function that passes through pre-selected continuum points. We note that identifying the continuum points is not trivial,  since the spectrum is densely populated with spectral lines. Hence, it is possible that the normalisation procedure introduces offsets to our model comparison. However, in light of the other uncertainties, this results in negligible errors.

We attempted to extract the visual spectra of the primary and secondary  using the technique of spectral disentangling \citep{Hadrava1995, Shenar2020, Shenar2022, Bodensteiner2020BeHR}. However, the technique results in multiple spurious features in the spectrum of the secondary that are the result of strong spectral-line variability observed in the primary. Somewhat surprisingly, no clear absorption features in the disentangled spectrum of the secondary are seen in lines belonging to the Balmer series. Our suspicion is that the Balmer absorption lines of the companion are masked by additional sources of emission in the system, or that an opaque disc hides the faint signatures of the secondary in the visual \citep{Koubsky2007}. Since the impact of the secondary is negligible in the visual, we measure the RVs of all HERMES spectra and add them in a common frame-of-reference for the subsequent analysis.

Additionally, we retrieved  UV spectra from the Mikulski Archive for Space Telescopes (MAST) acquired with the IUE in high-dispersion large aperture mode in the years 1978--1985 (Program IDs: PG2SS, HRDAK, IBHGM, HA191). The data are thoroughly described in \citet{Parthasarathy1986} and \citet{Dudley1990}. For the quantitative spectroscopy, we  use far UV spectra covering the spectral range 1150--1975\,\AA\ at a resolving power of $R\approx 10000$ and S/N$\approx 5$--$10$. However, we also use near-UV IUE spectra of a similar resolution covering the spectral range 1900--3125\,\AA\ to support the analysis of the spectral energy distribution (SED).

Finally, for the SED analysis, we also compile photometry \citep{Morel1978}.

% ==========================================================
\subsection{Model atmospheres}
\label{subsec:AtmospherePoWR}
% ==========================================================

We perform a quantitative spectroscopic analysis using the Potsdam Wolf Rayet (PoWR) code (\citealt*{PoWR2002}; \citealt{PoWR2003,PoWR2015}), which solves the non-LTE radiative transfer problem in spherical symmetry. Aside from the abundances, the models are set by a few fundamental parameters: the base effective temperature $T_*$, the bolometric luminosity $L$, the surface gravity $g_*$, and wind parameters such as the mass-loss rate $\dot{M}$, terminal wind speed $\varv_\infty$, and the clumping factor $D$, which describes the density ratio of clumps compared to an equivalent smooth wind. The radius $R_*$ is given by the Stefan-Boltzmann relation, $R_* \propto \sqrt(L)\,T_*^{-2}$. The effective temperature $T_*$ in PoWR is defined with respect to the layer in which the continuum Rosseland optical depth fulfills $\tau_{\rm Ross} = 20$. However, we also provide the fundamental parameters with respect to $\tau_{\rm Ross} = 2/3$, denoted here as $T_{\rm eff}, g, $ and $R$, since they are more useful when comparing to the evolution models. While $T_* \approx T_{\rm eff}$ for main sequence stars, they can diverge for stars with sparse atmospheres like that of the primary in $\upsilon$\,Sgr. In the subsonic regime, the density and velocity structures are set from (quasi)-hydrostatic equilibrium. In the supersonic regime, the velocity field is assumed to follow a $\beta$-law with $\beta = 0.8$ \citep*{CAK1975, Friend1986}. In the main non-LTE iteration, the line profiles are assumed to have a constant Doppler width, which is set to $30\,\kms$ for both stars. During the formal integration, the Doppler widths comprise of depth-dependent thermal broadening and microturbulence $\xi$, as described by \citet{Shenar2015}. 

\citet{Dudley1992} derived $\xi = 8\pm2\,\kms$ for the primary A-type supergiant. We computed models with values in the range $3 < \xi < 10 \,\kms$ and find that $\xi = 5\,\kms$ provides a better fit to the overall spectrum, though the results depend on the lines used. This may be partly due to usage of non-LTE models in our work. We therefore fix $\xi = 5\,\kms$, noting that this parameter impacts primarily the metallic abundances that are not fitted in our work. For the companion, which can only be seen in the far UV, we adopt $\xi = 20\,\kms$ for simplicity and faster computation, which has negligible impact on the derived physical parameters within errors. The microturbulence is assumed to grow in proportion to the wind velocity with a multiplicative fraction of 10\%. An attempt to derive the projected rotational velocity of the primary $\varv \sin i_1$ via Fourier analysis \citep{Gray1992, Simon-Diaz2007} did not yield consistent results between the different lines, likely since the line profiles are dominated by variable non-rotational broadening processes. The line profiles  can be modelled satisfactorily with a projected rotational velocity of $\varv_1 \sin i = 10\,\kms$ and a macroturbulent velocity of $\varv_{\rm turb} = 25\,\kms$ via convolution with rotational and radial-tangential line profiles \citep{Gray1992}, respectively, and so these values are adopted. We note, however, that the results depend on the lines used. 

In this study, we focus on the fundamental physical parameters of the two components. We therefore fix the clumping factor at $D=1$ for both components. For the primary star, we adopt $\log \dot{M} = -7.0\,[\smy]$, which has no significant impact on the synthetic spectrum. While diagnostic lines such as H$\alpha$ show strong emission \citep{Koubsky2006,Bonneau2011}, the origin of this emission (discs, streams, or radial outflows) is not clear, and we therefore refrain from deriving $\dot{M}_1$. Similarly, the wind speed of the primary is arbitrarily  set to $v_{\infty,1} = 500\,\kms$, having no impact on the analysis.

A helpful constraint for the analysis is the light ratio of the components in the visual, which was found to be $\Delta m= 3.59\pm0.16\,$mag at a central wavelength of 700\,nm from interferometry \citep{Hutter2021}. Since both components are in the Rayleigh-Jeans domain in the visual, we can assume $\Delta V = 3.59\pm0.16$, which implies that the primary contributes $96.4\pm0.5$\% to the visual flux. 

For the analysis of the primary, we computed a representative grid of PoWR models with varying values of $\log g_{*, 1}~(1.1$--$1.8\,[\cms]), T_{*, 1}~(8$--$13\,{\rm kK})$ and hydrogen mass fraction $0.00001 < X_{\rm H, 1} < 0.1$.  Although \citet{Dudley1992} provided a detailed derivation of the chemical abundances, we fix them to Solar here since the derivation was performed in the framework of an LTE analysis. The remaining elements are kept fixed to Solar according to \citep{Asplund2009}: $X_{\rm C} = 2.4\times 10^{-3}$, $X_{\rm N} = 6.9\times 10^{-4}$, $X_{\rm O} = 5.7\times 10^{-3}$, $X_{\rm Si} = 6.7\times 10^{-4}$, $X_{\rm S} = 3.1\times 10^{-4}$, $X_{\rm Mg} = 6.9\times 10^{-4}$, $X_{\rm Ne} = 1.3\times 10^{-3}$, $X_{\rm Fe} = 1.2\times 10^{-3}$, where the latter is  the mass fraction of the iron group elements (which includes mainly iron but also elements such as titanium and chromium).  For the secondary, which we argue is uncovered for the first time in our work, we use a published PoWR grid \citep{Hainich2019} to estimate its fundamental parameters, and recompute an adjusted model to fit the P-Cygni signatures (Sect.\,\ref{subsubsec:winds}).

The analysis of the A2~I primary in $\upsilon$\,Sgr is challenging for several reasons: (a) With its low gravity, the primary is at the edge of hydrostatic stability, which is reflected in convergence difficulties of the models. (b) The combination of low gravity and a relatively high temperature results in extreme non-LTE conditions, reflected in a non-linearity of the models. (c) Multiple analysis efforts in the past noted a discrepancy between various spectral diagnostics  of the temperature in the primary, making it thus far impossible to reproduce with standard spherically-symmetric  models \citep{Hack1963, Dudley1990}. (d) The spectrum is extremely dense in lines, causing severe line blending and making the normalisation of the spectra challenging. Given the highly non-linear behaviour of the models, contradictory diagnostics, and extensive computation time, we do not perform an automated or statistical fit, but rather demonstrate the impact of fundamental parameters on key diagnostic lines to estimate the physical properties of the system. Errors follow from a by-eye comparison of adjacent models, or from error propagation.

% ==========================================================
\subsection{Results}
\label{subsec:results}
% ==========================================================

\renewcommand{\arraystretch}{1.1}
%TTTTTTTTTTTTTTTTTTTTTTTTTTTTTTTTTTTTTTTTTTTTTTTTTTTTTTTTTTTTTT
\begin{table}
\centering
\caption{Derived parameters of $\upsilon$\,Sgr. Values of $T_*, g_*, R_*$ refer to $\tau_{\rm Ross} = 20$, while $T_{\rm eff}, g$, and $R$ refer to the photosphere ($\tau_{\rm Ross} = 2/3$)}
\begin{threeparttable}
\resizebox{.4\textwidth}{!}{\begin{tabular}{lc}
\hline
\hline
\vspace{-4mm}\\ 
Distance [pc]$^{\rm \, (a)}$  & $481_{-38}^{+46}$ \\
$f_1/f_{\rm tot} (V)^{\rm \,(b)}$ & $96.4\pm 0.5$\% \\
$T_{*, 1}\,$[kK]  & $10\pm1$  \\
$T_{\rm eff, 1}\,$[kK]  & $9\pm1$  \\
$T_{*, 2}\,$[kK]  & $23\pm2$  \\
$T_{\rm eff, 2}\,$[kK]  & $23\pm2$  \\
$X_{\rm H, 1}\,$  & $0.001 \pm 0.5$\,dex  \\
$E_{B - V}\,$[mag]  & $0.18\pm0.01$  \\
$\log L_{\rm 1}\,[\mathrm{L}_\odot]$   & $3.67 \pm 0.15$  \\
$\log L_{\rm 2}\,[\mathrm{L}_\odot]$  & $3.1\pm0.2$  \\
$\log g_{*, 1}\,[\cms]$   & $1.2_{-0.2}^{+0.1}$ \\
$\log g_1\,[\cms]$   & $1.0_{-0.2}^{+0.1}$ \\
$R_{*, 1}\,[\mathrm{R}_\odot]$  & $23^{+7}_{-6}$  \\
$R_{1}\,[\mathrm{R}_\odot]$  & $28^{+9}_{-7}$  \\
$R_{*, 2}\,[\mathrm{R}_\odot]$  & $2.2\pm 0.3$  \\
$R_{\rm 2}\,[\mathrm{R}_\odot]$  & $2.2\pm 0.3$  \\
$M_{\rm spec, 1}\,[\mathrm{M}_\odot]$  & $0.3_{-0.2}^{+0.5}$  \\
$M_{\rm 2, ev}\,[\mathrm{M}_\odot]^{\rm \, (c)}$  & $6.8\pm0.8$  \\
% $M_2 \sin^3 i\,[\mathrm{M}_\odot]$  & $2.0\pm 0.3$  \\
$i[^\circ]^{\rm \, (d)}$  & $50^{+10}_{-20}$  \\
% $M_2 [\mathrm{M}_\odot]$  & $4.4_{-2.0}^{+3.6}$  \\
$v \sin i_{\rm A}\,[\kms]$ & $\lesssim 10$    \\
$v \sin i_{\rm B}\,[\kms]$ & $250\pm20$    \\
$v_{\rm mac, 1}\,[\kms]$ & $\approx 20$--$40$       \\
\hline
\end{tabular}}
\footnotesize
\begin{tablenotes}
(a) \cite{Bailer-Jones2021}\\ (b) \cite{Hutter2021}\\ (c) Derived from $\log L_2$ and $T_{\rm eff, 2}$ using the BONNSAI tool \citep{Bonnsai2014}\\ (d) \cite{Netolicky2009}
\end{tablenotes}
\end{threeparttable}
\label{tab:Parameters}
\end{table}
%TTTTTTTTTTTTTTTTTTTTTTTTTTTTTTTTTTTTTTTTTTTTTTTTTTTTTTTTTTTTTT
The derived parameters for the system are provided in Table\,\ref{tab:Parameters}. A detailed account of the analysis procedure and our findings is given below.

\subsubsection{The A2~I Primary}
\label{subsubsec:priman}

\begin{figure}
   \centering
\includegraphics[width=.48\textwidth]{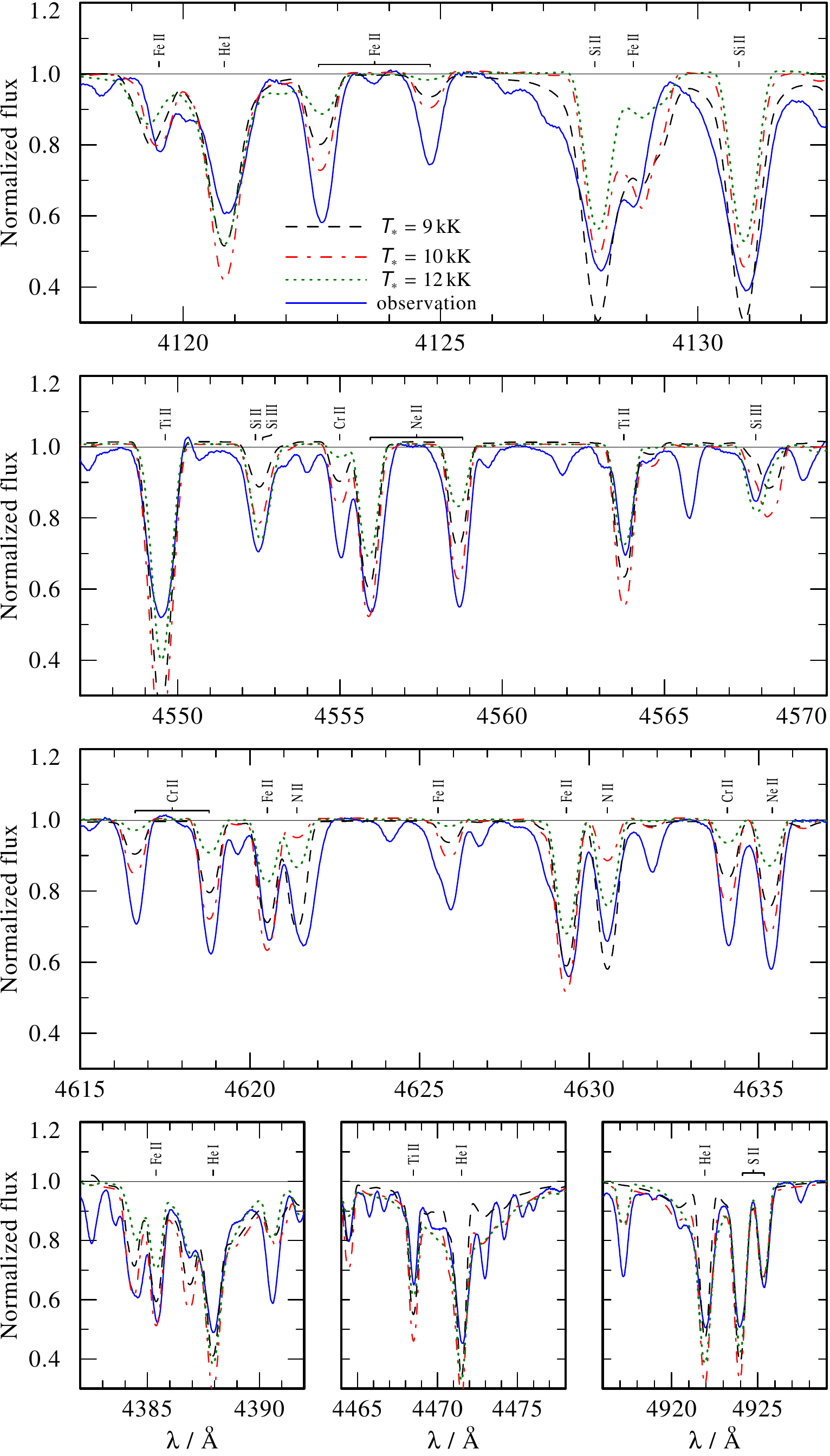}
    \caption{Comparison between the observed spectrum of multiple metal lines and He\,{\sc i} lines with three PoWR models with differing effective temperatures of $T_{*, 1} = 9, 10$, and $12\,$kK. The 12\,kK model has $\log g_* = 1.5\,[\cms]$, while the other two models have $\log g_* = 1.2\,[\cms]$, because convergence of lower $\log g_*$ values for the hotter model is unstable. The secondary (estimated to contribute less than 5\%) is neglected here.}
    \label{fig:Metals}
\end{figure}

\begin{figure}
   \centering
\includegraphics[width=.48\textwidth]{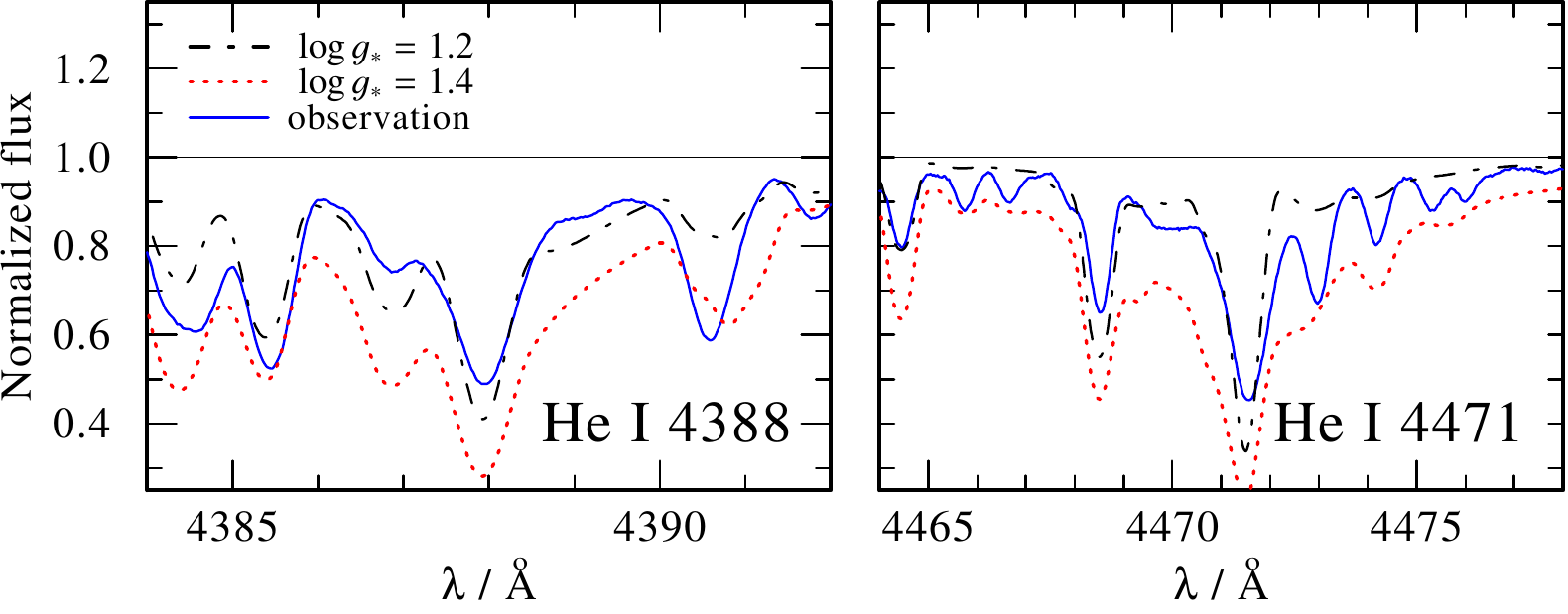}
    \caption{Comparison between the observed spectrum of two He\,{\sc i} lines and two models with $T_* = 9$\,kK and differing $\log g_*$ (1.2, 1.4\,$[\cms]$). The $9$\,kK model is shown for this illustration because the preferred model with 10\,kK overproduces the strength of the He\,{\sc i} lines (Fig.\,\ref{fig:Metals}). Regardless, the conclusions remain the same for the 10\,kK model. }
    \label{fig:Grav}
\end{figure}

A substantial simplification in the analysis arises from the fact that that secondary's contribution to the visual flux is negligible, and we therefore ignore it for the purpose of the analysis of the visual spectrum. The visual spectrum of $\upsilon$\,Sgr is dominated mainly by singly-ionised elements such as Si\,{\sc ii}, Fe\,{\sc ii}, Ti\,{\sc ii}, Ne\,{\sc ii}, and Cr\,{\sc ii}, though some lines belonging to neutral elements (e.g., Fe\,{\sc i}, He\,{\sc i}) and doubly-ionised elements (e.g., C\,{\sc iii}, Si\,{\sc iii}) are also present. 
It is extensively documented in the literature that  different diagnostic lines imply different temperatures, varying from values as low as 8\,kK from He\,{\sc i} lines and as high as 13.5\,kK from the Si\,{\sc ii/iii} balance \citep{Dudley1993}. While using non-LTE seems to remove some of this tension, a single model could not be found which reproduces the strengths of He\,{\sc i} lines along with the ionic balance of metal lines. Fig.\,\ref{fig:Metals} shows selected ranges of metal lines and He\,{\sc i} lines for models with varying temperatures in the range $T_* = 9$--$12$\,kK, with the other parameters comparable to those derived by \citet{Dudley1992}. Avoiding CNO lines, which may well deviate from baseline, the majority of metal lines are better reproduced for temperatures of the order of $9$--$10\,$kK, though some are suggestive of higher temperatures (e.g., Si\,{\sc iii}\,$\lambda 4568$). The He\,{\sc i} lines point at either temperatures of $\approx 12\,$kK or higher, or $8\,$kK or lower, which is inconsistent with the metal lines. We suspect that this is related to the highly complex and delicate atmospheric structure of the primary. We do not find a clear need to adopt non-Solar abundances for non-CNO metal lines (such as Ne, as suggested previously by \citealt{Leushin2000}). We note that $\log g_*$ can also impact the ionisation balance. However, the impact of $T_*$ on the metal lines in the relevant temperature regime is significantly stronger, and we therefore rely on these lines for the temperature estimate. The approximate errors provided should account for a possible $T_* - \log g_*$ degeneracy.

Despite some significant discrepancies, the $T_* = 10$\,kK model, which corresponds closely to $T_{\rm eff} = 9\,$kK, provides the best overall agreement with the spectrum, and furthermore agrees with our analysis of the SED (Sect.\,\ref{subsubsec:SED}). This is lower than the value of $T_{\rm eff} = 11.8$\,kK derived by \citet{Dudley1993} from an SED analysis, or 11.75\,kK derived by \citet{Dudley1992} from detailed spectroscopy. However, the former did not account for the contribution of the secondary (which is significant at wavelengths of $\lesssim 3000\,\AA$), while the latter did not account for non-LTE effects. We suspect that this is the main origin for the discrepancy of our results with \citet{Dudley1992} and \citet{Dudley1993}. 

With the temperature fixed, the gravity can be refined from the pressure-broadened profiles of H and He\,{\sc i} lines. However, because the strength of the hydrogen lines strongly depends on $X_{\rm H}$, and the stellar surface comprises almost entirely of helium, we use He\,{\sc i} lines for this purpose. Fig.\,\ref{fig:Grav} shows He\,{\sc i} lines for two distinct surface gravities, demonstrating that $\log g_* = 1.2\,[\cms]$ provides a good fit to the lines. Values larger than $\approx 1.3\,[\cms]$ or lower than $1.0\,[\cms]$ can be rejected. The corresponding photospheric value (i.e., at $\tau_{\rm Ross} = 2/3$) is $\log g = 1.0\,[\cms]$. We note that our derived $\log g_*$ depends on the fixed value of $T_*$. For example, a value of $T_* = 12\,$kK, which might be inferred from the He\,{\sc i} lines alone, would require $\log g_*= 1.5\,[\cms]$ to reproduce the wings of the He\,{\sc i} lines. However, as discussed above, a high-temperature solution is strongly disfavoured from the metal lines and the SED analysis. While such correlations could induce an additional error on $\log g$, we do not expect it to add significantly to the errors cited in Table\,\ref{tab:Parameters}.

\begin{figure}
   \centering
\includegraphics[width=.48\textwidth]{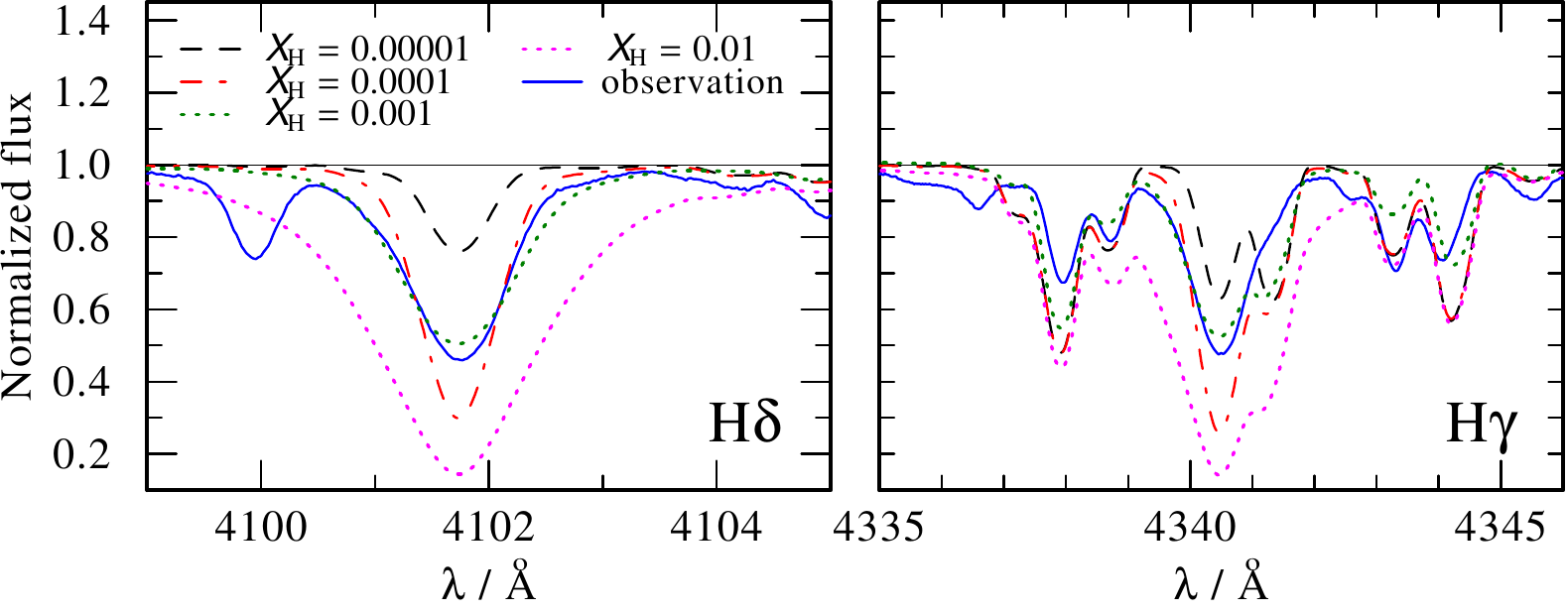}
    \caption{Shown are models with varying $X_{\rm H, 1}$ values and otherwise parameters identical to those given in Table\,\ref{tab:Parameters}.}
    \label{fig:Hyd}
\end{figure}

With $\log g$ and $T_{\rm eff}$ derived, we can advance with the derivation of the surface hydrogen mass fraction $X_{\rm H}$. The impact of changing $X_{\rm H}$ for various Balmer lines is shown in Fig.\,\ref{fig:Hyd}. We find a hydrogen mass fraction of $X_{\rm H} = 0.001$ to an accuracy of a factor of three roughly. Hence, the surface of the primary comprises almost solely of helium, as was already proposed in previous studies \citep{Hack1966, Dudley1992}. 

\begin{figure}
   \centering
\includegraphics[width=.48\textwidth]{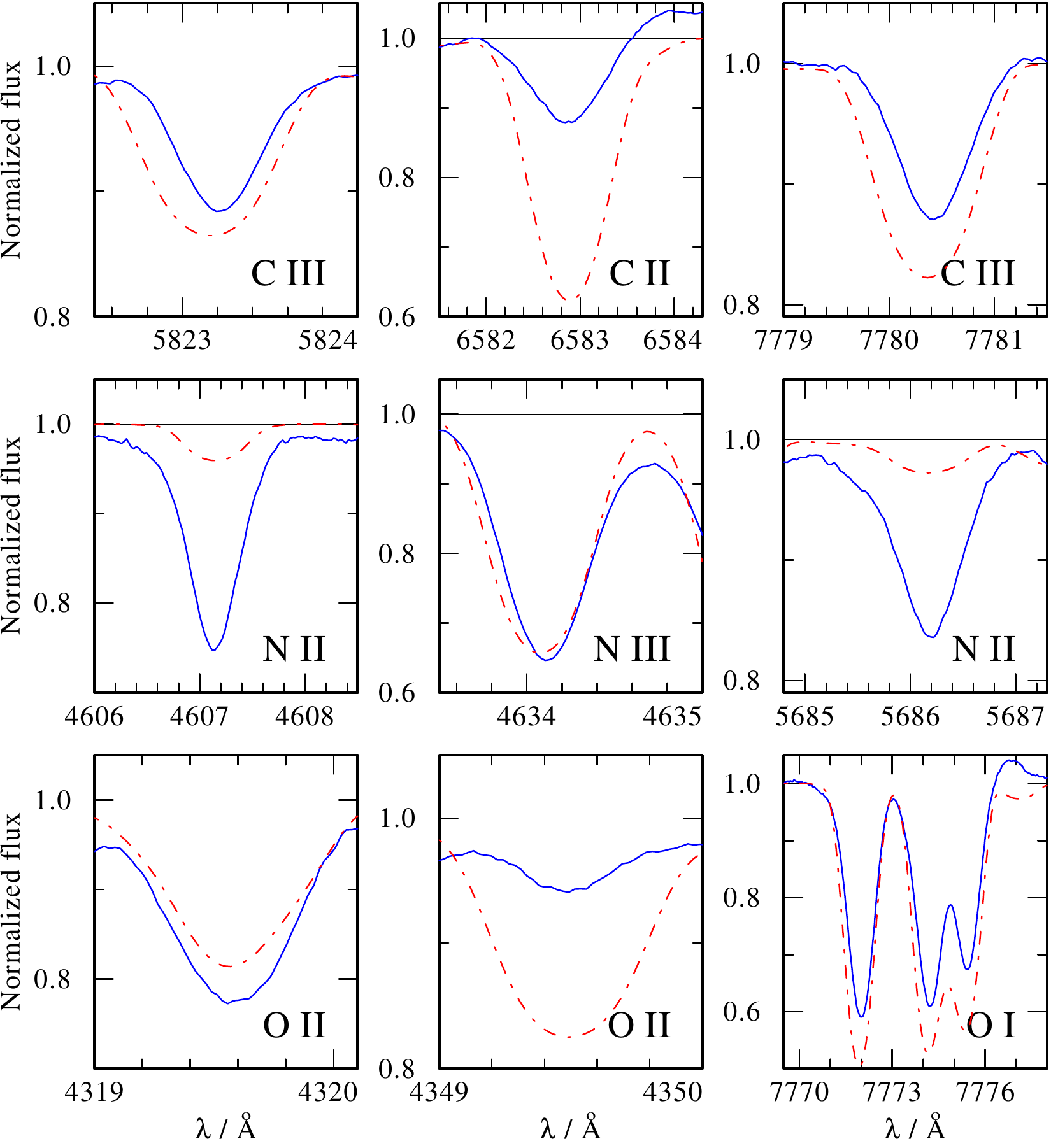}
    \caption{Comparison between our chosen model ($T_{*, 1} = 10\,$kK, $\log g_* = 1.2\,[\cms]$, $X_{\rm H} = 0.001$) and the observations, zooming-in on several CNO lines. The mismatch tentatively suggests an increased N and decreased CO abundance, but this is not seen consistently among all lines.}
    \label{fig:CNO}
\end{figure}

A comparison between our chosen representative model and several CNO lines is shown in Fig.\,\ref{fig:CNO}. The majority of nitrogen lines imply an enhanced nitrogen abundance by a factor of 4--5, while a few (such as N\,{\sc iii}\,$\lambda 4634$) seem to fit well with baseline abundance. Similarly, some C and O lines suggest a reduced abundance, while others do not. We note that the lines are extremely sensitive to the physical parameters of the star and that no model can satisfactorily reproduce their ionic ratios simultaneously to the metallic ones and the He\,{\sc i} lines. We therefore conclude that the spectrum contains tentative evidence for CNO processed material in the atmosphere of the primary.

\subsubsection{Analysis of the secondary and the SED}
\label{subsubsec:SED}

%FFFFFFFFFFFFFFFFFFFFFFFFFFFFFFFFFFFFFFFFFFFFFFFFFFFFFFFFFFFFFF
\begin{figure*}
   \centering
\includegraphics[width=\textwidth]{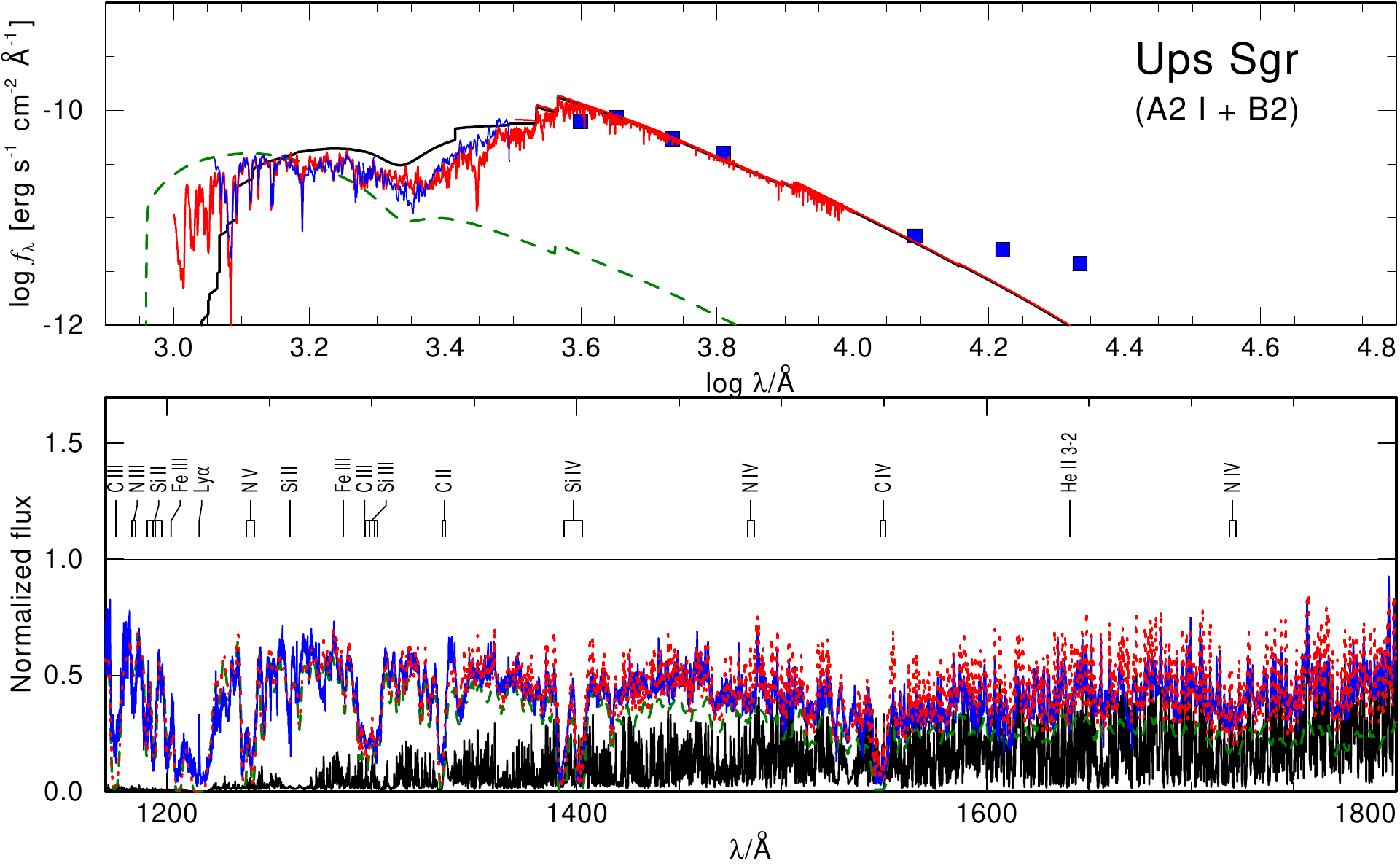}
    \caption{\textit{Upper panel:} Comparison between the observed IUE co-added spectrum (blue line) and photometry and our PoWR model for $\upsilon$\,Sgr. The model (red) comprises the models of the cooler primary (black) and hotter secondary (green). The composite model includes both continua and line transitions, but only continua are shown for the component SEDs for clarity. \textit{Lower panel:} Zoom-in on the match between the IUE spectrum and the composite PoWR model, showing also the component spectra. The IUE spectrum is normalised using the model continuum. Evidently, the secondary fully dominates the spectrum bluewards of $\approx 1500\,$\AA. The excess seen in the infrared likely originates in the disc observed in the system \citep{Netolicky2009}. }
    \label{fig:SED}
\end{figure*}
%FFFFFFFFFFFFFFFFFFFFFFFFFFFFFFFFFFFFFFFFFFFFFFFFFFFFFFFFFFFFFF

%FFFFFFFFFFFFFFFFFFFFFFFFFFFFFFFFFFFFFFFFFFFFFFFFFFFFFFFFFFFFFF
\begin{figure}
   \centering
\includegraphics[width=.48\textwidth]{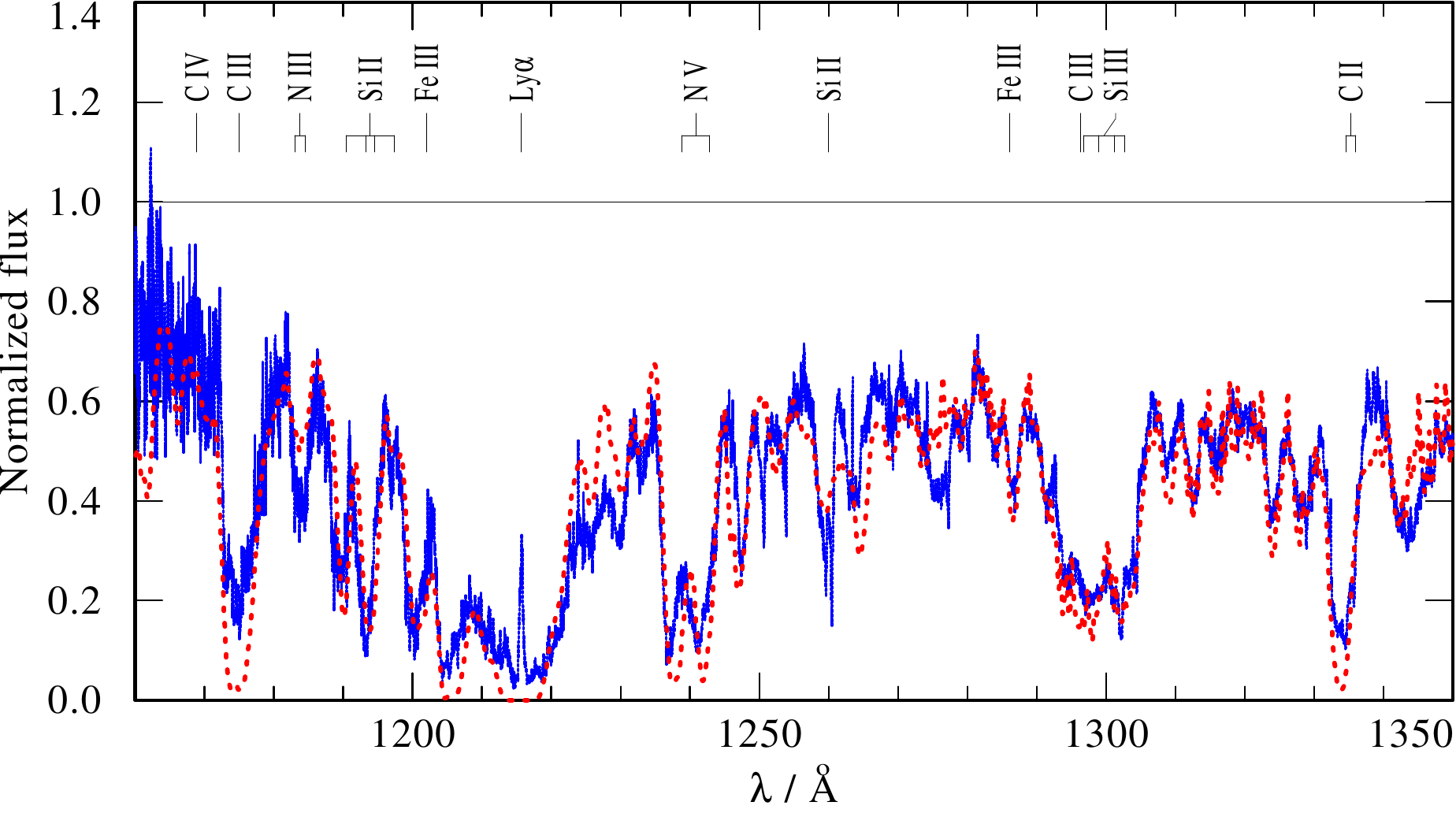}
    \caption{Zoom-in on the UV region dominated by the companion. The presence of a rapidly rotating hot star is evident from the spectrum, in contrast to previous reports of a slowly rotating star in the system \citep{Dudley1990}.} 
    \label{fig:ZoomComp}
\end{figure}
%FFFFFFFFFFFFFFFFFFFFFFFFFFFFFFFFFFFFFFFFFFFFFFFFFFFFFFFFFFFFFF

We advance to the far UV and SED modelling of the system. Previous attempts to model the SED of $\upsilon$\,Sgr accounted only for the primary \citep{Dudley1993}. However, as it turns out, the companion contributes roughly 50\% of the flux in the near-UV, a contribution that increases to over 90\% in the far UV. Upon inspection of the UV, it becomes readily clear that it is dominated by a different source. First, the strong signature of the iron forest of the primary, which appears as noise but is actually of astrophysical origin, gradually decreases bluewards of $\approx 1500\,\AA$. Then, instead of the sharp iron-forest lines, one observes broad and round stellar lines belonging to ions such as Si\,{\sc iii} and Fe\,{\sc iii} (Figs.\,\ref{fig:SED} and \ref{fig:ZoomComp}). These clearly originate in a companion. The spectral features imply temperatures in the range $20$--$25\,$kK, which match roughly a B2~V star.

To model the spectrum and the SED, we combine the  PoWR model of the primary with a model computed for the secondary source with $T_{*, 2} \approx T_{\rm eff, 2} = 23\,$kK. Lacking diagnostics for the gravity,  we fix $\log g_{*, 2} = 4.0\,[\cms]$ (the default value of the PoWR grids). We use the law derived by \cite*{Cardelli1989} for the reddening with a total-to-selective extinction ratio of $R_V = 3.1$. The result is shown in Fig.\,\ref{fig:SED}. We find an extinction of $E_{B - V} = 0.18\,$mag and luminosities of $\log L_1 = 3.67\pm0.15\,[\mathrm{L}_\odot]$ and $\log L_2 = 3.1\pm0.2\,[\mathrm{L}_\odot]$ for the adopted Gaia distance of $d = 481^{+46}_{-38}\,$pc \citep{Bailer-Jones2021}. A zoom-in on the blue part of the IUE spectrum, which is dominated by the secondary, is shown in Fig.\,\ref{fig:ZoomComp}. By fitting the line profiles, we estimate $\varv \sin i_2 = 250\pm20\,\kms$.  The broad and round line profiles of the secondary stand in stark contrasts to the results obtained by \citet{Dudley1990}, who extracted the spectrum of the secondary through a spectral deconvolution technique (similar to disentangling). Their extracted spectrum is nearly flat with the exception of a few features which have the width of a few pixels. As the presence of a rapidly rotating companion is readily seen in the spectrum, we cannot but conclude that the measurements performed by \citet{Dudley1990} are spurious. The reason for the presence of these spurious features could be an interplay between instrumental effects and the extraction method implemented by \citet{Dudley1990}, such as bad pixels, albeit we could not reproduce this behaviour.

%FFFFFFFFFFFFFFFFFFFFFFFFFFFFFFFFFFFFFFFFFFFFFFFFFFFFFFFFFFFFFF
\begin{figure}
   \centering
\includegraphics[width=.48\textwidth]{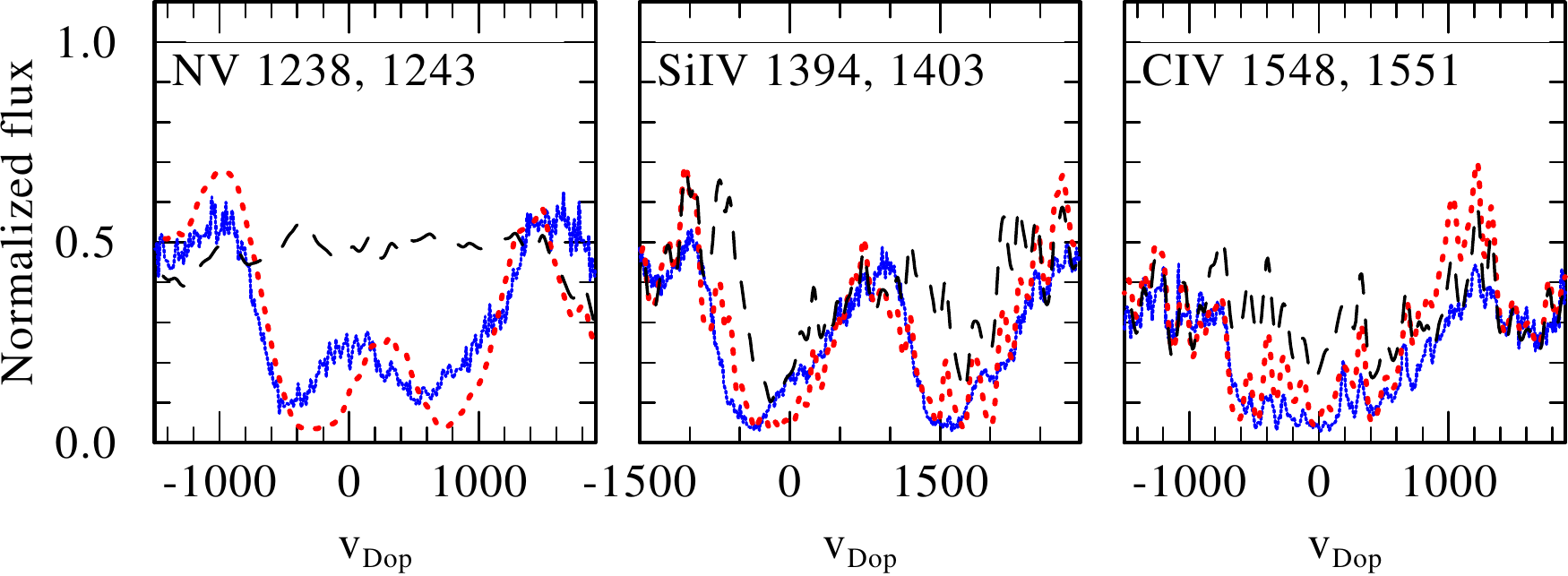}
    \caption{Zoom-in on the P-Cygni resonance lines in the UV, showing the observations (blue) compared to the composite PoWR model, which is dominated by the secondary, once without additional X-rays (black dashed line) and once with superionisation by X-rays (red dotted line).
    } 
    \label{fig:Wind}
\end{figure}
%FFFFFFFFFFFFFFFFFFFFFFFFFFFFFFFFFFFFFFFFFFFFFFFFFFFFFFFFFFFFFF

Aside from broad photospheric features, the UV spectrum also contains strong P-Cygni lines associated with the high-ionisation resonance lines N\,{\sc v}\,$\lambda \lambda 1238, 1243$, Si\,{\sc iv}\,$\lambda \lambda 1294, 1403$, and C\,{\sc iv}\,$\lambda \lambda 1548, 1551$. An attempt to model these lines as an outflow originating in the secondary star was performed by \citet{Dudley1992}, though not with a consistent atmosphere model. We find that these lines can indeed be relatively well reproduced with a wind model for the secondary (Fig.\,\ref{fig:Wind}), and we obtain a good fit with $\log \dot{M}_2 = -7.0\,[\smy]$ and $\varv_{\infty, 2} =300\,\kms$. However, these features may also originate in other forms of outflow (e.g., jets), and the wind properties provided here should therefore only be considered as illustrative.

To populate high ionisation stages such as N\,{\sc v}, Auger ionisation via an additional X-ray source was invoked \citep{Cassinelli1979}. X-rays are commonly observed in early-type stars \citep*[e.g.][]{Oskinova2006}, presumably owing to the presence of shocks in their winds \citep*{Feldmeier1997}. Alternatively, in this system the X-rays may also originate from the impact of outflowing material from the primary to the secondary. The modelling of X-rays in PoWR relies on various free parameters (X-ray temperature, onset radius, filling factor), and a description can be found in \citet{Baum1992} and \citet{Shenar2015}. Generally, the impact of X-rays is akin to a ``switch'': if sufficient X-ray photons of high enough energies are present, high ionisation levels will be populated. The X-ray parameters are such that the model of the secondary has a modest X-ray luminosity of $L_{X} = 10^{32}\,{\rm erg}\,{\rm s}^{-1}$. 

\subsubsection{Revised orbital analysis}
\label{subsubsec:orbit}

\begin{figure}
   \centering
\includegraphics[width=.48\textwidth]{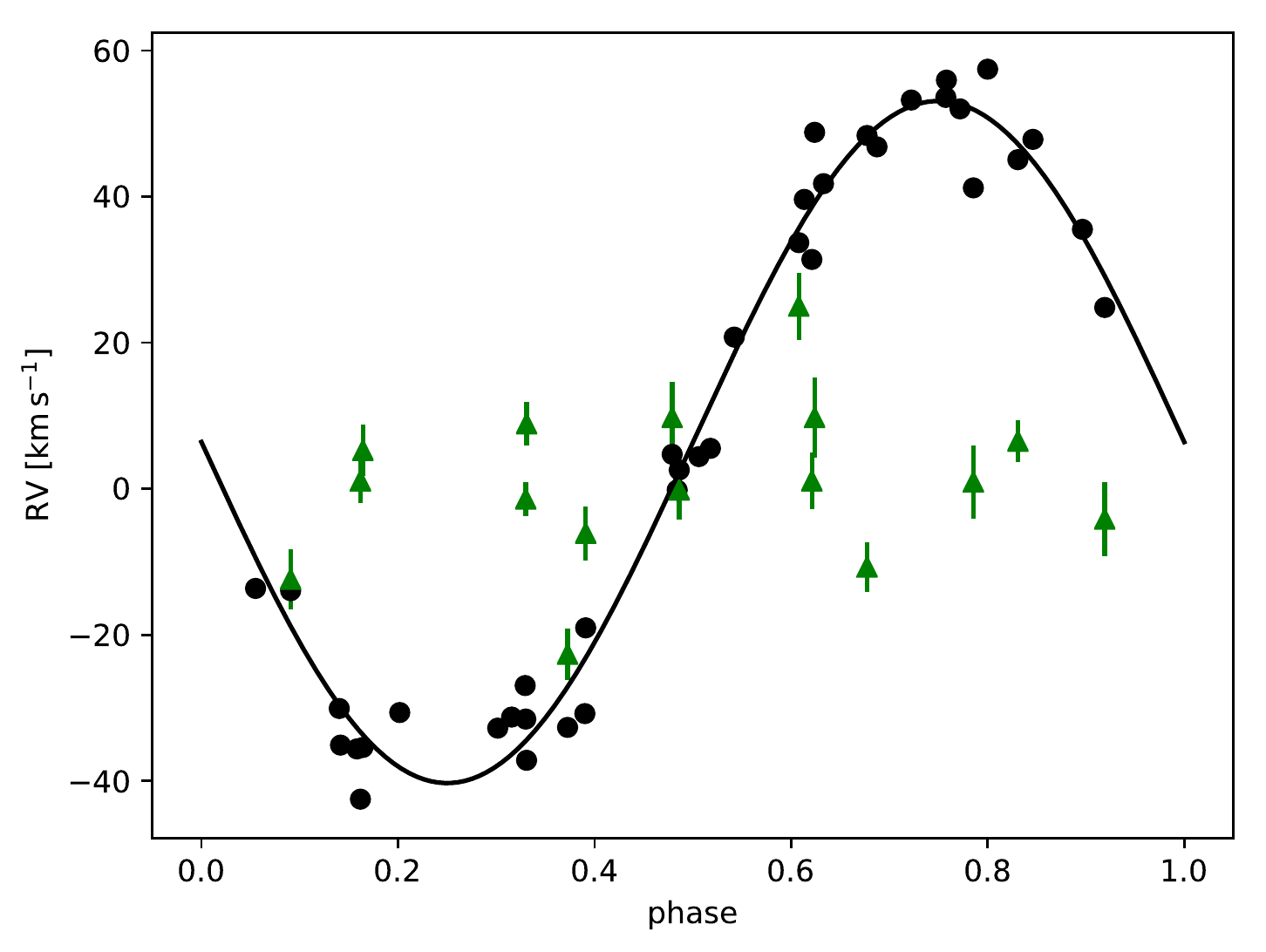}
    \caption{Orbital solution to the RVs of both components of $\upsilon$\,Sgr. The secondary's RVs are poorly determined, but yield a better fit for low amplitudes ($K_2 \lesssim 5\,\kms$), and can rule out the high RV semi-amplitude of $\approx 30\,\kms$ derived by \citet{Dudley1990}.} 
    \label{fig:RVs}
\end{figure}

We measured the RVs of the primary by cross-correlating the HERMES spectra with a template \citep{Zucker1994}. First, we use our PoWR model as a template. Then, we co-add all observations in the same frame-of-reference, and use the co-added spectrum to remeasure the RVs \citep{Shenar2017b, Dsilva2020}. We use several ranges that are densely populated with spectral lines, and find that the choice does not impact the results. As the system is known to have a circular orbit, we fix the eccentricity to $e=0$ and the argument of periastron $\omega = 90^\circ$, such that phase 0 corresponds to the time in which the primary is in front of the secondary. The period is also fixed to $P = 137.9343\,$d \citep{Koubsky2006}. The RVs are fit to an orbital solution to obtain the time of periastron $T_0$, RV semi-amplitude $K_1$, and systemic velocity $V_0$ using the Python \texttt{lmfit}\footnote{\href{https://lmfit.github.io/lmfit-py/}{https://lmfit.github.io/lmfit-py/}} package \citep{Newville2014}. The results are shown in Fig.\,\ref{fig:RVs}, and imply $K_1 = 46.7 \pm 0.8\,\kms$, in good agreement with previous estimates \citep[e.g.][]{Koubsky2006}.

We attempted to measure the RVs of the secondary by repeating  cross-correlating our spectral template with the far-UV data. However, the lines are too broad, and the S/N too low, to obtain meaningful results. The RVs scatter irregularly at a high amplitude, but an orbital fitting through them yields small amplitudes of the order of 5\,$\kms$. Even though we cannot robustly establish the value of $K_2$, the measurements imply that $K_2 \lesssim 10\,\kms$. This is in stark contrast to the high amplitude ($\approx 30\,\kms$) derived by \citet{Dudley1993}.

%FFFFFFFFFFFFFFFFFFFFFFFFFFFFFFFFFFFFFFFFFFFFFFFFFFFFFFFFFFFFFF
%FFFFFFFFFFFFFFFFFFFFFFFFFFFFFFFFFFFFFFFFFFFFFFFFFFFFFFFFFFFFFF

% ==========================================================
\section{Stellar evolution models}
\label{sec:StellarEvolution}
% ==========================================================

To explain the nature of the $\upsilon$\,Sgr system and its evolutionary history we compute stellar evolution tracks as detailed below.

% ==========================================================
\subsection{Method and assumptions}
\label{subsec:MESAmethod}
% ==========================================================

We use the Modules for Experiments in Stellar Astrophysics code (\textsc{mesa}, version 15140, \citealt{Paxton2011,Paxton2013,Paxton2015,Paxton2018,Paxton2019}) to evolve non-rotating stellar models. The metallicity is $Z=0.014$. Both stars are evolved until the surface gravity of the primary reaches $\log g = 7$, indicating its transition into a white dwarf.

% ==========================================================
\subsubsection{Microphysics}
\label{subsubsec:mesa}
% ==========================================================

The equation of state employed by \textsc{mesa} is a blend between OPAL \citep{Rogers2002}, SCVH \citep{Saumon1995}, FreeEOS \citep{Irwin2004}, HELM \citep{Timmes2000}, and PC \citep{Potekhin2010}.

Radiative opacities are primarily from OPAL \citep{Iglesias1993, Iglesias1996}, with low-temperature data from \citet{Ferguson2005} and the high-temperature, Compton-scattering dominated regime according to \citet{Buchler1976}. Electron conduction opacities follow \citet{Cassisi2007}.

We use the built-in \textsc{mesa} nuclear reaction network \texttt{approx21}. The Joint Institute for Nuclear Astrophysics REACLIB reaction rates \citep{Cyburt2010} are used, with additional tabulated weak reaction rates (\citealt*{Fuller1985}; \citealt{Oda1994, Langanke2000}) and screening via the prescriptions of \citet{Salpeter1954}, \citet*{Dewitt1973}, \citet{Alastuey1978} and \citet{Itoh1979}. Thermal neutrino loss rates follow the formulae of \citet{Itoh1996}.

% ==========================================================
\subsubsection{Mixing}
\label{subsubsec:mixing}
% ==========================================================

The Ledoux stability criterion is used to define convective regions, where mixing is treated according to mixing-length theory \citep*[MLT;][]{MLT1958,MLT} with a mixing-length parameter of $\alpha_\mathrm{MLT}=2$. Overshooting above the hydrogen-burning convective core during the MS evolution follows the exponentially decaying prescription of \cite{Herwig2000}, with a decay scale of $f_\mathrm{ov} H_P$, where  $f_\mathrm{ov}=0.016$ and $H_P$ is the pressure scale height. We do not apply overshooting to later stages or convective shells, and do not employ the MLT++ treatment of \textsc{mesa} for superadiabatic convection \citep{Paxton2013}.

% ==========================================================
\subsubsection{Mass transfer}
\label{subsubsec:masstransfer}
% ==========================================================

The mass-transfer rate during RLOF is computed according to \cite{Kolb1990}. Roche-lobe radii are computed using the fit provided by \cite{Eggleton1983}. We employ a prescription which continuously updates the mass-transfer efficiency during the stellar evolution computation, following \cite{GilkisVinkEldridgeTout2019}. When the mass transfer is not conservative, the material lost from the system is assumed to carry away the specific angular momentum at the accretor.

% ==========================================================
\subsubsection{Wind mass loss}
\label{subsubsec:winds}
% ==========================================================

For hot (effective surface temperatures of $T_\mathrm{eff} \ge 11000\,\mathrm{K}$) hydrogen-rich ($X_\mathrm{s} \ge 0.7$) phases of the evolution, wind mass loss follows prescription provided by \cite{VS21}, which is based on the works of \cite*{Vink1999,Vink2000,Vink2001}. For $X_\mathrm{s} \le 0.4$ we use the theoretical recipe provided by \cite{Vink2017}, and in the range $0.4 < X_\mathrm{s} < 0.7$ we interpolate between \cite{VS21} and \cite{Vink2017}. For cool ($T_\mathrm{eff} \le 10000\,\mathrm{K}$) phases of the evolution, the empirical relation given by \cite*{deJager1988} is employed. For $10000\,\mathrm{K} < T_\mathrm{eff} < 11000\,\mathrm{K}$ the wind mass-loss rate is interpolated between the hot and cool prescriptions.

% ==========================================================
\subsection{Evolutionary tracks and best-fitting model}
\label{subsec:tracks}
% ==========================================================

We construct evolutionary grids in two steps, first for a broad range of initial conditions, and then focusing on achieving the observed orbital period of $\upsilon$\,Sgr. The initial primary masses used are $M_\mathrm{ZAMS,1} = 4.25$, $4.5$, $4.75$, $5$, $5.25$, and $5.5\,\mathrm{M}_\odot$, the mass ratios $q=0.5$, $0.625$ and $0.75$, and initial periods $P_\mathrm{i} = 3$, $4$, $5$, $7$, $10$, $13$ and $18\,\mathrm{d}$. For each combination of $M_\mathrm{ZAMS,1}$ and $q$ we use the tracks computed in the first step to derive the initial period that will result in a final period of $P\simeq 138\,\mathrm{d}$.

%FFFFFFFFFFFFFFFFFFFFFFFFFFFFFFFFFFFFFFFFFFFFFFFFFFFFFFFFFFFFFF
\begin{figure*}
  \centering
    \begin{subfigure}{\textwidth}
    \centering
    \includegraphics[trim=0 115 0 0,clip,width=\textwidth]{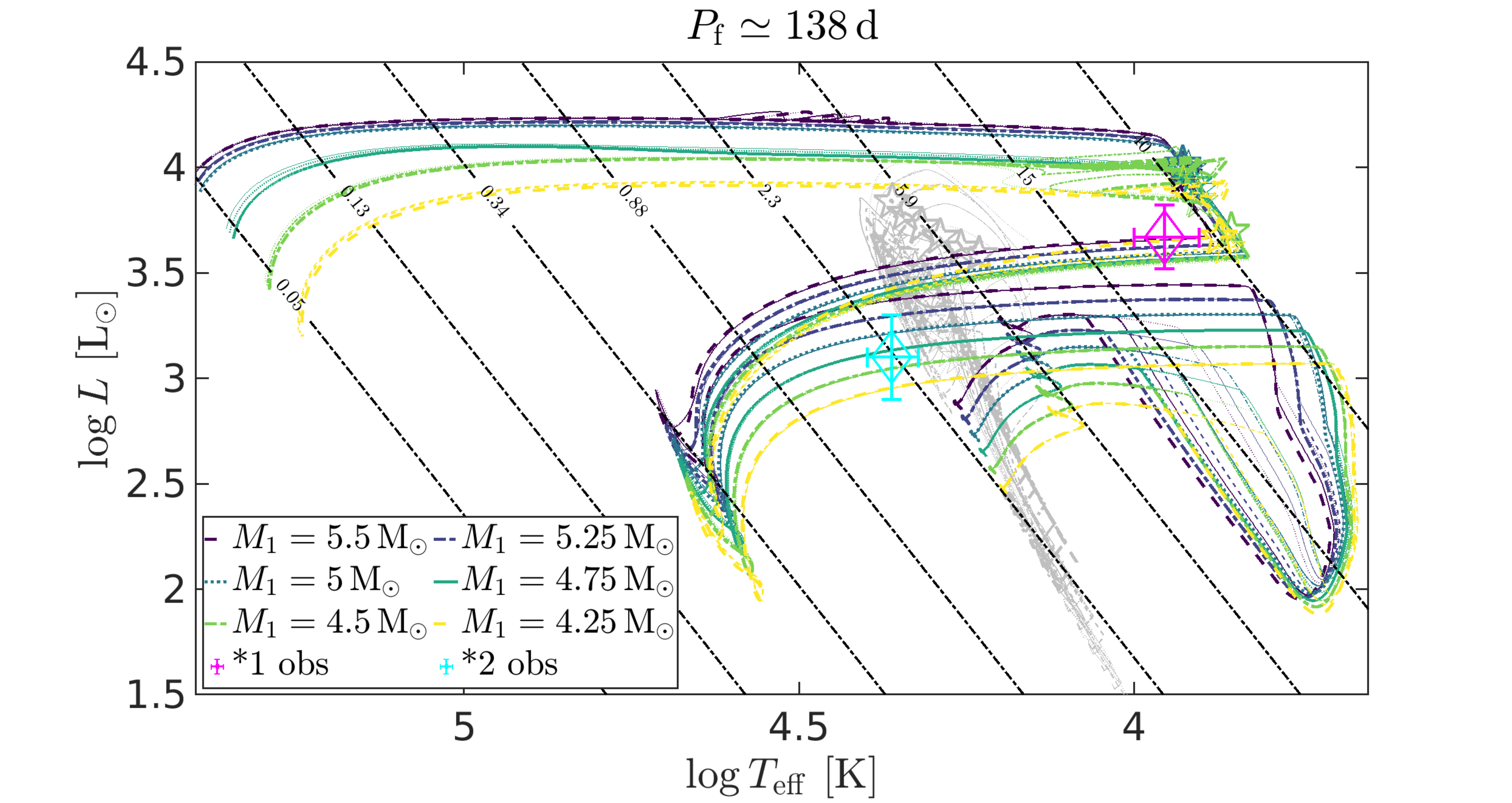}
    \label{fig:HRD}
  \end{subfigure}\\
  \begin{subfigure}{\textwidth}
    \centering
    \includegraphics[trim=0 0 0 75,clip,width=\textwidth]{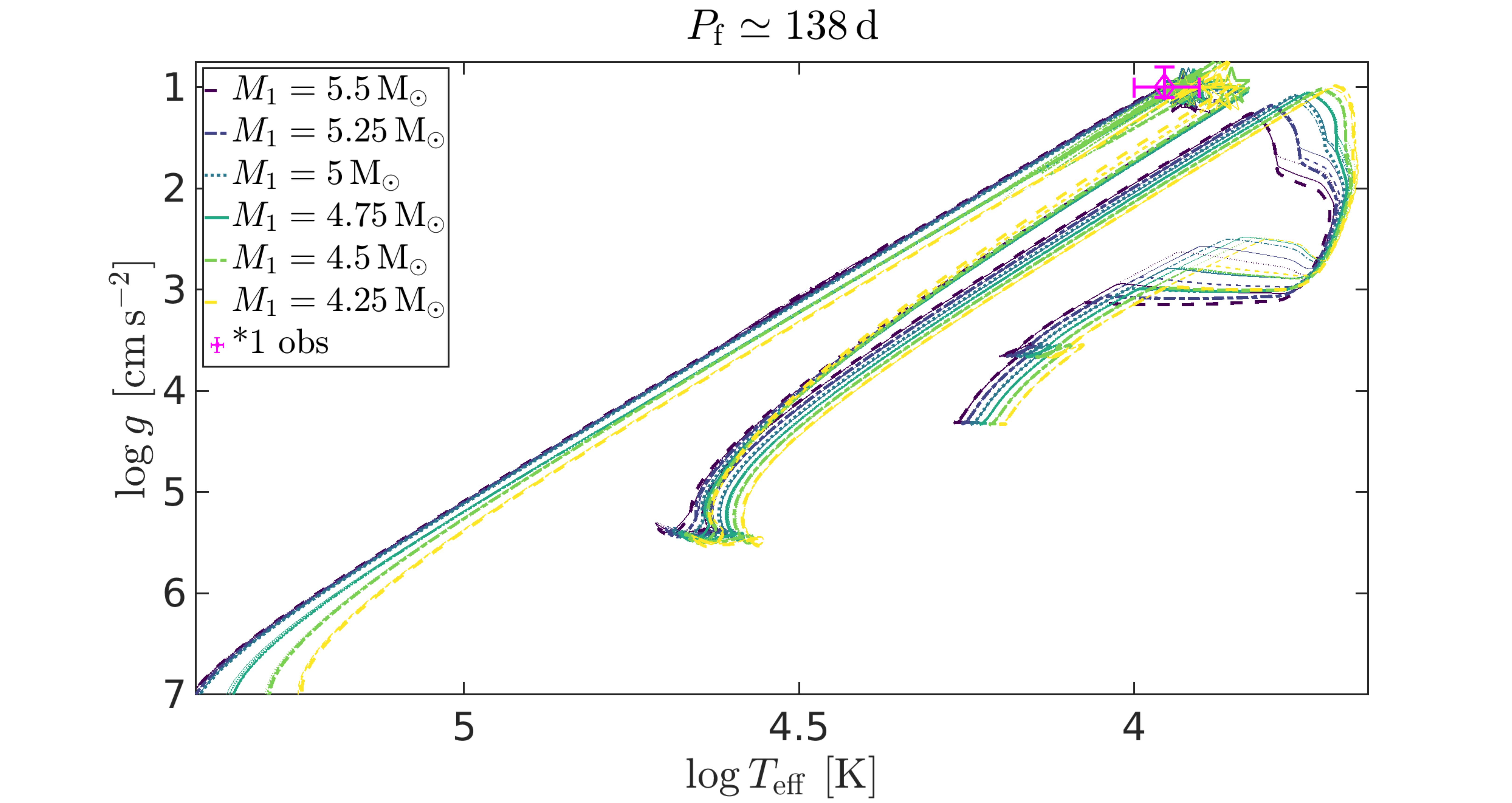}
    \label{fig:g_vs_T}
  \end{subfigure}\\    
  \caption{Evolutionary tracks simulated with \textsc{mesa} with initial masses colour-coded as shown in the inset. For each initial mass, three tracks are shown, for the initial companion masses and orbital periods as indicated in Table\,\ref{tab:masses}. \textit{Upper panel:} Hertzsprung-Russell diagram comparing the simulated tracks with the spectroscopically-derived parameters for both components of $\upsilon$\,Sgr. The evolution of the primary is coloured by its initial mass, and the evolution of the companion is shown in grey. The point on the evolutionary track that minimises $\chi^2$ (Eq.\,\ref{eq:chi2}) is marked with a star symbol. Lines of constant stellar radius are labeled for each $R$ shown in units of $\mathrm{R}_\odot$. \textit{Lower panel:} Surface gravity as function of effective surface temperature for the primary in the \textsc{mesa} evolutionary tracks compared to the spectroscopic fit.} 
  \label{fig:tracks}
\end{figure*}
%FFFFFFFFFFFFFFFFFFFFFFFFFFFFFFFFFFFFFFFFFFFFFFFFFFFFFFFFFFFFFF
%TTTTTTTTTTTTTTTTTTTTTTTTTTTTTTTTTTTTTTTTTTTTTTTTTTTTTTTTTTTTTT
\begin{table}
\centering
\caption{Initial conditions for stellar evolution calculations.}
\begin{threeparttable}
\begin{tabular}{ccc}
\hline
$M_1/ \mathrm{M}_\odot$ & $M_2/ \mathrm{M}_\odot$ &  $P_\mathrm{i} / \mathrm{d}$ \\
\hline
$4.25$ & $2.125$  & $13.089$ \\
$4.25$ & $2.6562$  & $8.452$ \\
$4.25$ & $3.1875$  & $ 6.025$ \\
\hline
$4.5$ & $2.25$  & $13.57$ \\
$4.5$ & $2.8125$  & $8.156$ \\
$4.5$ & $3.375$ & $6.027$ \\
\hline
$4.75$ & $2.375$  & $13.797$ \\
$4.75$ & $2.9688$ & $8.298$ \\
$4.75$ & $3.5625$  & $5.988$ \\
\hline
$5$ & $2.5$  & $13.438$ \\
$5$ & $3.125$ & $8.4$ \\
$5$ & $3.75$  & $6.045$ \\
\hline
$5.25$ & $2.625$  & $12.005$ \\
$5.25$ & $3.2812$ & $7.272$ \\
$5.25$ & $3.9375$  & $5.355$ \\
\hline
$5.5$ & $2.75$  & $10.908$ \\
$5.5$ & $3.4375$ & $6.595$ \\
$5.5$ & $4.125$  & $4.872$ \\
\hline
\hline
\end{tabular}
\footnotesize
\end{threeparttable}
\label{tab:masses}
\end{table}
%TTTTTTTTTTTTTTTTTTTTTTTTTTTTTTTTTTTTTTTTTTTTTTTTTTTTTTTTTTTTTT
In Fig.\,\ref{fig:tracks} we present $18$ stellar evolution tracks with initial conditions as listed in Table\,\ref{tab:masses}. Along each track we search for the best-fitting model according to $\chi^2$, computed as
\begin{multline}
    \chi^2 = \left( \frac{T_\mathrm{eff,1,obs} - T_\mathrm{eff,1,calc}}{\Delta T_\mathrm{eff,1,obs}}\right)^2
    + \left( \frac{\log L_\mathrm{1,obs} - \log L_\mathrm{1,calc}}{\Delta \log L_\mathrm{1,obs}}\right)^2 \\
    + \left( \frac{\log g_\mathrm{1,obs} - \log g_\mathrm{1,calc}}{\Delta \log g_\mathrm{1,obs}}\right)^2
    + \left( \frac{\log X_\mathrm{H,1,obs} - \log X_\mathrm{H,1,calc}}{\Delta \log X_\mathrm{H,1,obs}}\right)^2 \\
    + \left( \frac{T_\mathrm{eff,2,obs} - T_\mathrm{eff,2,calc}}{\Delta T_\mathrm{eff,2,obs}}\right)^2
    + \left( \frac{\log L_\mathrm{2,obs} - \log L_\mathrm{2,calc}}{\Delta \log L_\mathrm{2,obs}}\right)^2 \\
    + \left( \frac{P_\mathrm{orb,obs} - P_\mathrm{orb,calc}}{\Delta P_\mathrm{orb,obs}}\right)^2 ,
\label{eq:chi2}    
\end{multline}
with observed parameters and their errors taken from Table\,\ref{tab:Parameters}. While the orbital period has been measured very accurately, we take $\Delta P_\mathrm{orb,obs}=1$ to avoid giving an extremely high weight to the period in finding the best-fitting model. It can be seen that the best-fitting models are somewhat larger than the primary radius inferred from the spectroscopic analysis, clustering around $R\approx 40\,\mathrm{R}_\odot$. While a couple of tracks with $M_\mathrm{ZAMS,1}=4.5\,\mathrm{M}_\odot$ seem to be closest to the spectroscopic radius, the surface hydrogen mass fraction in these models is too high ($X_\mathrm{H,1} \ga 0.1$).

%FFFFFFFFFFFFFFFFFFFFFFFFFFFFFFFFFFFFFFFFFFFFFFFFFFFFFFFFFFFFFF
\begin{figure*}
   \centering
\includegraphics[width=\textwidth]{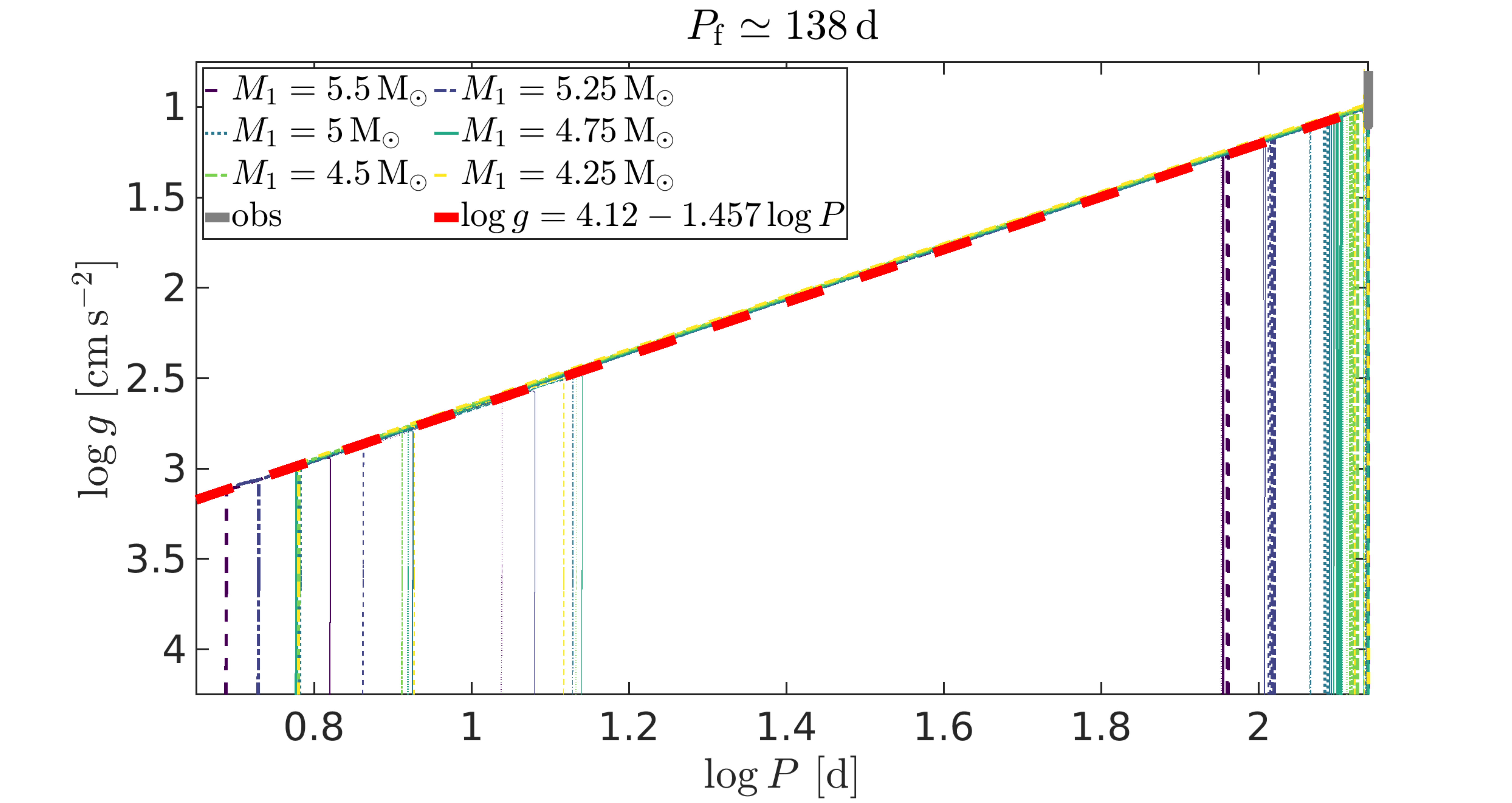}
    \caption{Surface gravity of the primary as function of the binary orbital period. Evolutionary phases appearing as vertical lines occur when there is no mass transfer and the period stays essentially constant, while during RLOF the evolution proceeds along the fitted $g\left(P\right)$ relation marked with the thick dashed red line. The region defined by the observed orbital period and spectroscopically-derived surface gravity is marked by a thick grey line.}
    \label{fig:g_vs_P}
\end{figure*}
%FFFFFFFFFFFFFFFFFFFFFFFFFFFFFFFFFFFFFFFFFFFFFFFFFFFFFFFFFFFFFF
In Fig.\,\ref{fig:g_vs_P} we show the evolution of the surface gravity of the primary as function of the binary orbital period. When there is no mass transfer the orbital period stays constant (mass loss by winds is negligible) while the primary expands or contracts. During RLOF the orbital period changes. We find that there is a well-defined relation between the surface gravity of the donor star and the binary orbital period, as shown in Fig.\,\ref{fig:g_vs_P}. This relation can be intuitively understood by considering the case of fully-conservative mass transfer and approximating
\begin{equation}
R_1\approx R_{\mathrm{RL},1} \approx \frac{2 a}{3^{4/3}} \left(\frac{M_1}{M_1+M_2}\right)^{1/3},
\label{eq:approxRL}
\end{equation}
with the approximation for $R_{\mathrm{RL},1}$ from \citealt{Paczynski1971a} accurate to within $2\%$ (for $M_1<0.8 M_2$) which yields
\begin{equation}
d \log g \approx - \left( \frac{4}{3} + \frac{1}{9} \frac{M_2}{M_2-M_1} \right) d \log P.
\label{eq:magic_g_P_relation}
\end{equation}
The slope in our fit (Fig.\,\ref{fig:g_vs_P}) is obtained by setting $M_2 = 10 M_1$ in Eq.\,\ref{eq:magic_g_P_relation}. While this is only an approximation, it explains most of the mass transfer phases, as the mass ratio is quickly inverted after a rapid initial mass transfer, and then proceeds with $M_2 \gg M_1$ and close to fully-conservative mass transfer. The late evolution is approximately governed by the behaviour described by Eq.\,\ref{eq:magic_g_P_relation}. Using the fitted relation, the observed orbital period gives $\log g \simeq 1$, independently of the spectroscopic analysis.

%FFFFFFFFFFFFFFFFFFFFFFFFFFFFFFFFFFFFFFFFFFFFFFFFFFFFFFFFFFFFFF
\begin{figure}
   \centering
\includegraphics[width=0.48\textwidth]{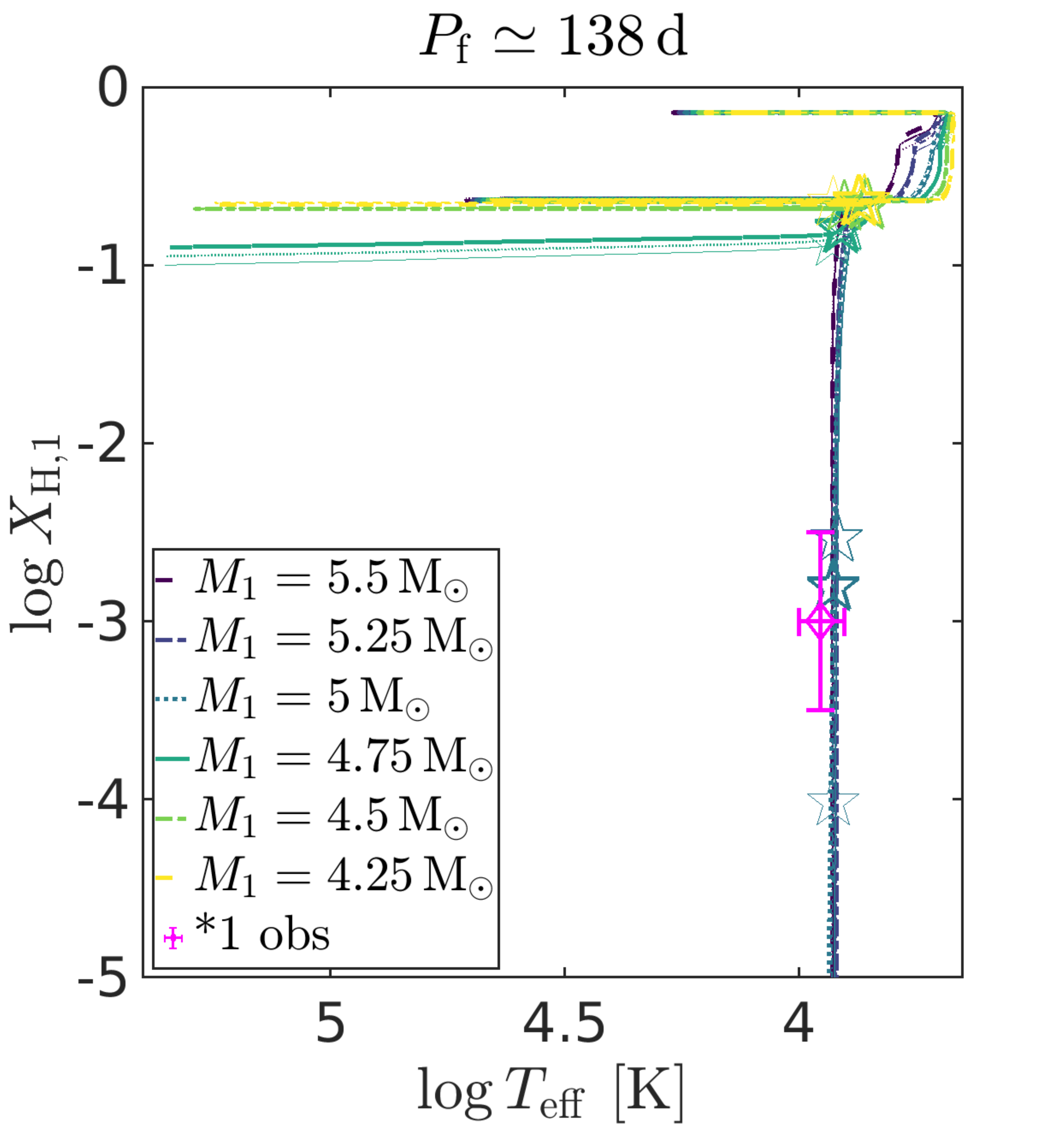}
    \caption{Surface hydrogen mass fraction of the primary as function of the effective surface temperature compared to the spectroscopically-derived parameters. The point that fits best the observations along each track is marked with a star symbol.} 
    \label{fig:XH_vs_Teff}
\end{figure}
%FFFFFFFFFFFFFFFFFFFFFFFFFFFFFFFFFFFFFFFFFFFFFFFFFFFFFFFFFFFFFF
A key property of $\upsilon$\,Sgr is its extreme hydrogen deficiency. In Fig.\,\ref{fig:XH_vs_Teff} we show the surface hydrogen mass fraction as function of effective surface temperature in our evolutionary tracks. We find that models with an initial mass of $< 5\,\mathrm{M}_\odot$ stay hydrogenous ($X_\mathrm{H,1} > 0.1$) until the end of their evolution, and therefore cannot explain $\upsilon$\,Sgr. This can be understood as the helium shell-burning phase gives rise to a large radial expansion only for massive enough helium stars (\citealt*{Jeffery1988, Hurley2000}; Appendix\,\ref{sec:appendixa}), and corresponding carbon-oxygen cores, and low-mass helium stars contract to become white dwarfs without completely being depleted of hydrogen. For $M_\mathrm{ZAMS,1} \ge 5\,\mathrm{M}_\odot$, the second mass transfer stage results in drastic depletion of the envelope hydrogen. With increasing mass the hydrogen depletion rate becomes faster, and therefore it is reasonable that the best-fitting model has the minimal mass that eventually results in complete hydrogen depletion, as it spends a relatively long time in the mass-transfer phase.

%TTTTTTTTTTTTTTTTTTTTTTTTTTTTTTTTTTTTTTTTTTTTTTTTTTTTTTTTTTTTTT
\begin{table}
\centering
\caption{Parameters of best-fitting \textsc{mesa} model for $\upsilon$\,Sgr.}
\begin{threeparttable}
\resizebox{.4\textwidth}{!}{\begin{tabular}{lc}
\hline
\hline
\vspace{-4mm}\\ 
Age & $120\,\mathrm{Myr}$ \\
$T_{\rm eff, 1}\,$[kK]  & $8.3$  \\
$T_{\rm eff, 2}\,$[kK]  & $21.2$  \\
$X_{\rm H, 1}\,$  & $0.003$  \\
$X_{\rm He, 1}\,$  & $0.9835$  \\
$X_{\rm C, 1}\,$  & $7\times 10^{-5}$  \\
$X_{\rm N, 1}\,$  & $0.008$  \\
$X_{\rm O, 1}\,$  & $4\times 10^{-4}$  \\
$\log L_{\rm 1}\,[\mathrm{L}_\odot]$   & $4.0$  \\
$\log L_{\rm 2}\,[\mathrm{L}_\odot]$  & $3.5$  \\
$\log g_1\,[\cms]$   & $1.0$  \\
$\log g_2\,[\cms]$  & $4.0$  \\
$R_{\rm RL , 1}\,[R_\odot]$  & $48.9$  \\
$R_{\rm 1}\,[\mathrm{R}_\odot]$  & $47.8$  \\
$R_{\rm 2}\,[\mathrm{R}_\odot]$  & $4.2$  \\
$M_1\,[\mathrm{M}_\odot]$  & $0.84$  \\
$M_2\,[\mathrm{M}_\odot]$  & $7.25$  \\
$\dot{M}_{\rm RLOF,1}\,[\mathrm{M}_\odot\,\mathrm{yr}^{-1}]$ & $3.3\times 10^{-7}$\\
$\dot{M}_{\rm wind,1}\,[\mathrm{M}_\odot\,\mathrm{yr}^{-1}]$ & $2.2\times 10^{-8}$\\
\hline
\end{tabular}}
\footnotesize
\end{threeparttable}
\label{tab:ParametersMESA}
\end{table}
%TTTTTTTTTTTTTTTTTTTTTTTTTTTTTTTTTTTTTTTTTTTTTTTTTTTTTTTTTTTTTT
%FFFFFFFFFFFFFFFFFFFFFFFFFFFFFFFFFFFFFFFFFFFFFFFFFFFFFFFFFFFFFF
\begin{figure*}
    \includegraphics[width=\textwidth]{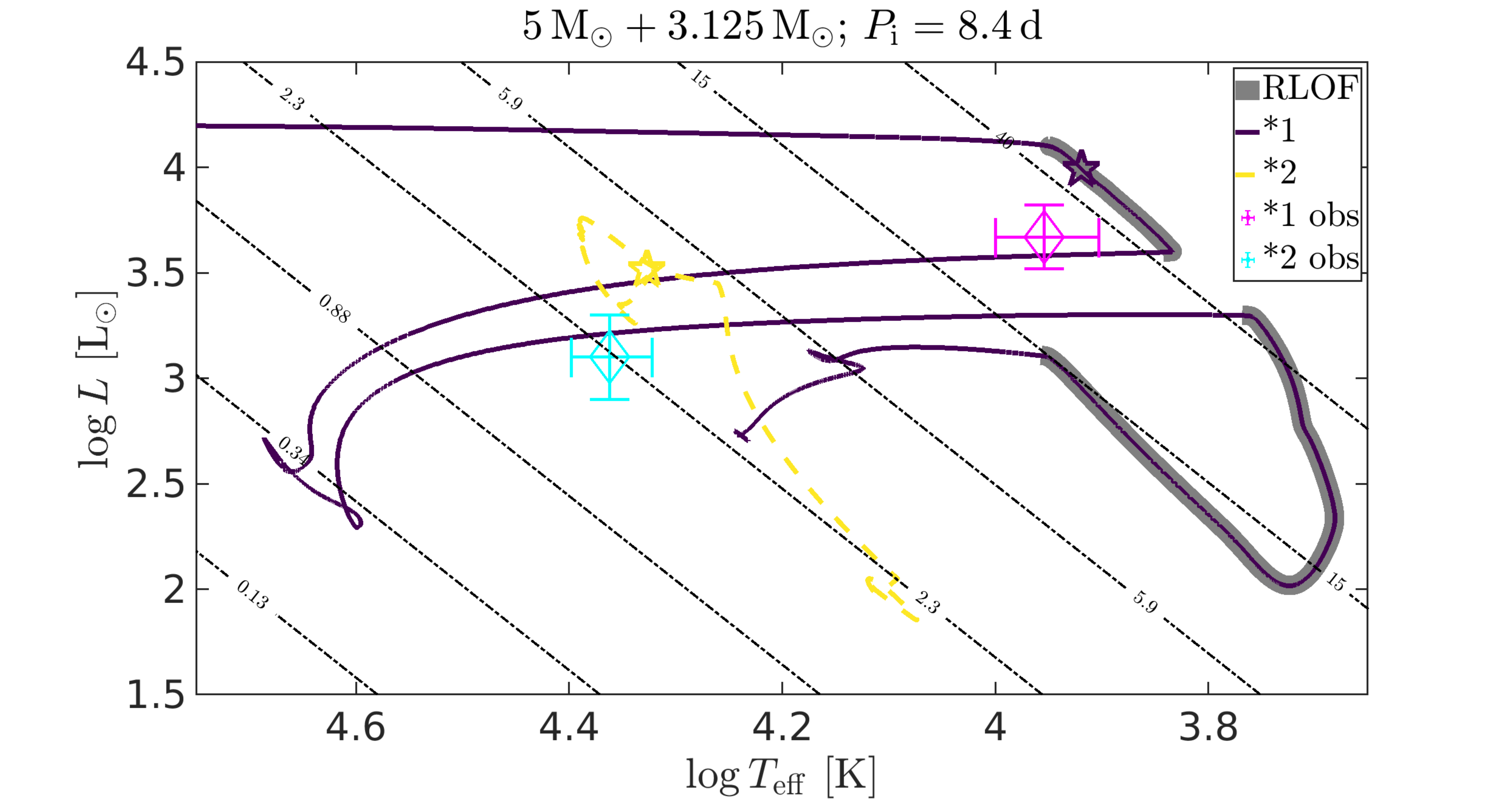}
  \caption{Evolution of the binary system that best matches the observed parameters of $\upsilon$\,Sgr, with the best-fitting model of the two components marked with star symbols, compared with the spectroscopically-derived parameters. Phases of mass transfer by RLOF are highlighted in grey. Lines of constant stellar radius are labeled for each $R$ shown in units of $\mathrm{R}_\odot$.}
  \label{fig:EvolBest}
\end{figure*}
%FFFFFFFFFFFFFFFFFFFFFFFFFFFFFFFFFFFFFFFFFFFFFFFFFFFFFFFFFFFFFF
%FFFFFFFFFFFFFFFFFFFFFFFFFFFFFFFFFFFFFFFFFFFFFFFFFFFFFFFFFFFFFF
\begin{figure}
    \includegraphics[width=0.48\textwidth]{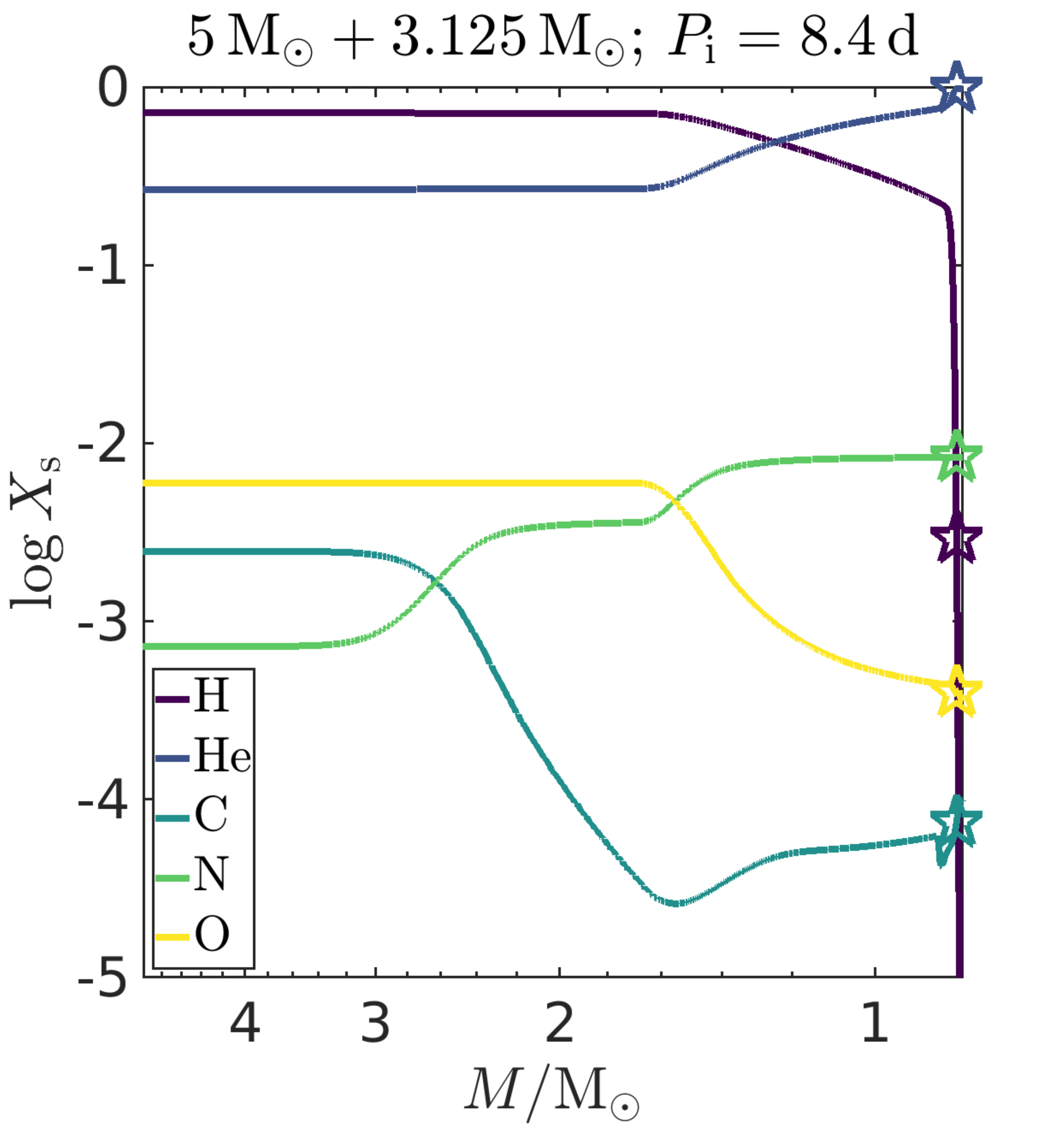}
  \caption{Surface composition as function of the total stellar mass of the primary in the evolution track that best matches $\upsilon$\,Sgr. The best-fit point in the evolution is marked with star symbols, when the primary mass has reduced to $M_1=0.84\,\mathrm{M}_\odot$.}
  \label{fig:XsBest}
\end{figure}
%FFFFFFFFFFFFFFFFFFFFFFFFFFFFFFFFFFFFFFFFFFFFFFFFFFFFFFFFFFFFFF
The best-fitting models according to the $\chi^2$ computation are for $M_\mathrm{ZAMS,1}=5\,\mathrm{M}_\odot$, and the lowest $\chi^2$ is achieved for the track with $M_\mathrm{ZAMS,2}=3.125\,\mathrm{M}_\odot$ and $P_\mathrm{i}=8.4\,\mathrm{d}$, whose evolution is shown in Fig.\,\ref{fig:EvolBest}, and parameters detailed in Table\,\ref{tab:ParametersMESA}. The evolution of the surface composition of our best-fitting model is presented in Fig.\,\ref{fig:XsBest}, showing the depletion of hydrogen and the enhancement of nitrogen.

% ==========================================================
\section{Discussion}
\label{sec:discussion}
% ==========================================================

\subsection{The mass of the primary}
\label{subsec:DiscussM1}

From the derived values of $T_{\rm eff, 1}$ and $L_1$, the Stephan-Boltzmann law implies $R_1 = 29\pm8\,\mathrm{R}_\odot$. This is almost a factor three lower than the value derived by \citet{Dudley1992} and \citet{Dudley1993}, which is mainly due to their adopted distance of 1500\,pc (instead of 481\,pc adopted here from Gaia). Using the derived surface gravity $\log g_1$ and radius $R_1$ for the primary, we find its spectroscopically-derived mass to be $M_1 = 0.3_{-0.2}^{+0.5}\,\mathrm{M}_\odot$. This is almost a factor 10 lower than proposed by \citet{Dudley1990}, who derived the minimum mass of the star from their measured kinematics of the system. While the uncertainties in the spectroscopic mass are substantial, a mass exceeding the Chandrasekhar limit can be ruled out at more than 95\% confidence.

As illustrated in Sect.\,\ref{subsubsec:orbit}, the $K_2$ semi-amplitude of $29.7\pm1.7$ derived by \citet{Dudley1990} cannot be confirmed here. The non-sinusoidal and rather erratic scatter of the RVs seen in Fig.\,\ref{fig:RVs} suggests that they are dominated by errors, but if the scatter were due to orbital motion, the peak-to-peak variability between the first and second orbital segments is about $20\,\kms$, suggesting a semi-amplitude of $10\,\kms$ or lower. In fact, our results point to an RV semi-amplitude that is as low as $K_2 \approx 3\,\kms$, though our measurement errors and data quality do not allow us to obtain a meaningful measurement for $K_2$. Hence, we can conclude that the mass ratio and minimum masses derived by \citet{Dudley1990} are spurious. 

One may wonder what exactly it is that \citet{Dudley1990} measured. Since we could not recover their measurements, we cannot fully answer this. However, a glance at their figure 4, which shows the extracted spectra of the primary and secondary, reveals that the narrow features in the spectrum of the secondary have narrow emission counterparts in the spectrum of the primary. This strongly suggests a cross-contamination between the two components.

The best-fitting model obtained from our \textsc{mesa} simulations has a mass of $M_1=0.84\,\mathrm{M}_\odot$ (Table\,\ref{tab:ParametersMESA}). This is higher than the derived value from spectroscopy ($0.3^{+0.5}_{-0.2}\,\mathrm{M}_\odot$), but consistent within 1$\sigma$. It is  still significantly  lower than the minimum mass computed by \citet{Dudley1990}, and also lower than the Chandrasekhar mass. Our analysis therefore rules out the possibility that $\upsilon$\,Sgr is an immediate core-collapse supernova progenitor.

\subsection{The radius of the primary}
\label{subsec:DiscussR1}

\begin{figure}
    \includegraphics[width=0.48\textwidth]{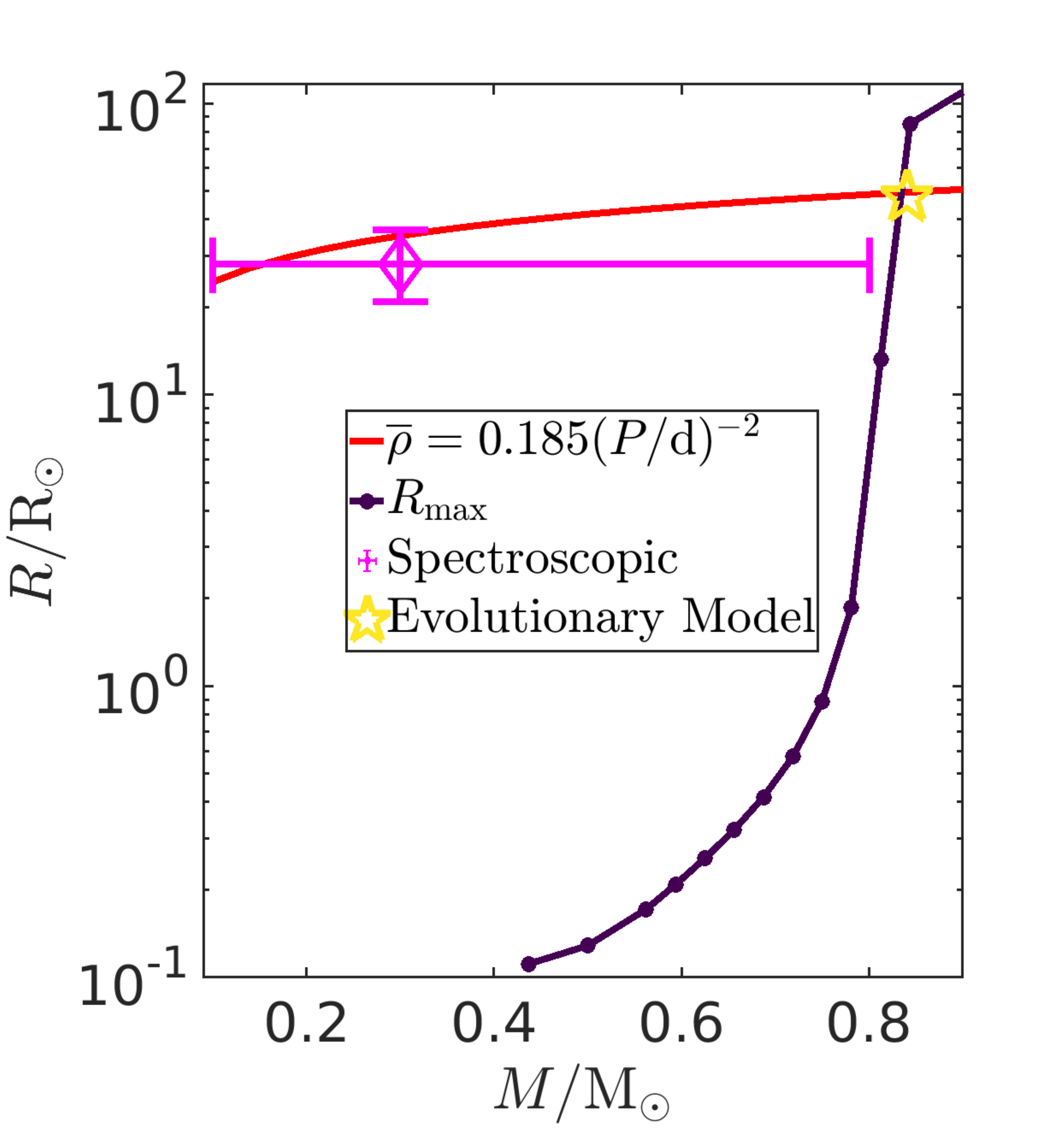}
  \caption{Stellar radius as function of mass according to Eq.~\ref{eq:Density} compared with single helium-star models and the evolutionary and spectroscopically-derived parameters.}
  \label{fig:RvsM}
\end{figure}

%The newly-derived mass of the primary is consistent with the fact that it is a Roche-lobe filling system. 
Is $\upsilon$\,Sgr a Roche-lobe filling system? \citet{El-Badry2022} showed that the average density of a Roche-lobe filling star fulfils to good approximation the following condition:
\begin{equation}
    \overline{\rho} = 0.185\,(P/{\rm d})^{-2}\,\,{\rm g}\,{\rm cm}^{-3},
\label{eq:Density}    
\end{equation}
which, for $\upsilon$\,Sgr, amounts to $\overline{\rho} = 9.86 \cdot 10^{-6}\,{\rm g}\,{\rm cm}^{-3}$. In Fig.\,\ref{fig:RvsM} we show the mass-radius relation obtained from $\overline{\rho}$ compared to our spectroscopically-derived mass and radius and best-fitting evolutionary model. In addition, we plot the maximum radius of single helium-star models (Appendix\,\ref{sec:appendixa}). Our best-fitting evolutionary model appears where the mass-radius relation using $\overline{\rho}$ from Eq.\,(\ref{eq:Density}) and the single helium-star models intersect. This can explain the low surface hydrogen mass fraction, as the primary is in the process of being stripped of its hydrogen and has almost become a helium star. Helium stars with lower masses will be much smaller, and cannot explain the properties of $\upsilon$\,Sgr.

We find tension between the spectroscopically-derived mass and radius and the best-fitting evolutionary model. This is also apparent in the Hertzsprung-Russell diagram (Fig.\,\ref{fig:EvolBest}). There are several possible explanations.
\begin{enumerate}
    \item There is a deficiency in the stellar evolution models. This is unlikely, as we have checked several assumptions in single helium-star models (Appendix\,\ref{sec:appendixa}) and our results are in general agreement with previous theoretical studies of helium stars \citep{Paczynski1971b,Jeffery1988,Hurley2000}.
    \item There is a deficiency in the model atmospheres. As thoroughly discussed in Sect.\,\ref{sec:specan}, the primary of  $\upsilon$\,Sgr is found in an extreme non-LTE regime with a very sparse atmosphere, and a  satisfactory model that fully reproduces the wealth of spectral lines cannot be obtained. While the errors are expected to encompass this difficulty, it is hard to rule out that systematics bias our results. On top of this, our analysis assumes the light ratio of the components as obtained from interferometry. However, if this measurement is biased by other sources (e.g., optically thin or optically thick discs), then this will have an impact on our analysis that is not considered in the errors. 
    \item The distance is underestimated. A distance that is $\approx 50\%$ larger would remove the tension in the radius and mass of the primary obtained from the spectroscopic and evolutionary analyses. The fact that $\upsilon$\,Sgr is a relatively wide binary, with the low-mass primary fully dominating the visual light, can mean that the Gaia distance (obtained from a single-star solution) is biased. While a binary solution was not published for $\upsilon\,$Sgr \citep{GaiaDR3}, the Gaia decision tree for publishing orbits was extremely conservative. We note, however, that the renormalised Unit Weight Error (RUWE) for $\upsilon$\,Sgr is 1.06. This is below the 1.4 threshold considered to be the upper limit for trustworthy solutions, and hence a clear indication that the distance is underestimated is lacking.
\end{enumerate}

%FFFFFFFFFFFFFFFFFFFFFFFFFFFFFFFFFFFFFFFFFFFFFFFFFFFFFFFFFFFFFF
%FFFFFFFFFFFFFFFFFFFFFFFFFFFFFFFFFFFFFFFFFFFFFFFFFFFFFFFFFFFFFF

\subsection{The nature of the companion}
\label{subsec:DiscussM2}

The faint companion in the $\upsilon$\,Sgr system is revealed to be a rapidly rotating B-type star (Fig.\,\ref{fig:ZoomComp}). For our estimated rotation velocity, the companion is spinning between $50\%$ and $90\%$ of its critical rotation rate, depending on the inclination. The MS equivalent that matches the luminosity and temperature of the companion in our spectral model (Fig.\,\ref{fig:SED}, Table\,\ref{tab:Parameters}) is a star with $M_\mathrm{ZAMS}\approx 6\,\mathrm{M}_\odot$. This is significantly higher than our estimated initial mass for the primary, and indicates that the companion has gained a large amount of mass.

We consider the possibility that some of the UV excess originates from energy released during accretion. Estimating the available accretion energy as $L_\mathrm{acc} \le G M_2 \dot{M}_\mathrm{RLOF} / R_2$ and adopting the companion parameters from Table\,\ref{tab:ParametersMESA} we get $L_\mathrm{acc}\la 18\,\mathrm{L}_\odot$, much lower than the stellar luminosity. Because of the relatively low mass-transfer rate at the observed state of the system, we conclude that any UV excess from the accretion process is negligible.

In our evolutionary model the mass of the companion has more than doubled, from $\approx 3\,\mathrm{M}_\odot$ to $\approx 7\,\mathrm{M}_\odot$. However, this result depends on our modelling assumptions, for example the efficiency of mass transfer. For less efficient mass transfer than we assume, the best-fitting model might have a higher initial mass and a smaller mass gain. Still, even for the most extreme possibility of nearly-equal initial masses the companion would have to gain $\ga 20\%$ of its ZAMS mass. This is at odds with the notion that accretion stops after a small increase in mass because of the accretor approaching critical rotation \citep{Packet1981,Renzo2021}, and might hint at a gap in our understanding of accretion in binaries.

% ==========================================================
\section{Summary and conclusions}
\label{sec:summary}
% ==========================================================

We have scrutinised the spectroscopic binary system $\upsilon$\,Sgr, providing a spectral model that explains its peculiar properties and reveals the rapidly rotating faint companion. The primary is a hydrogen-deficient ($X_\mathrm{H,1}\approx 0.001$) supergiant, with a low surface gravity ($\log g \approx 1.0$ [$\,\mathrm{cm}\,\mathrm{s}^{-2}$]). Using binary evolution models computed with \textsc{mesa}, we find a model that explains all these properties and matches the observed orbital period, in which the primary is a Roche-lobe filling evolved stripped helium star with a current mass of $M_1=0.84\,\mathrm{M}_\odot$ observed in the process of losing the remainder of its hydrogen.

We have the following conclusions:
\begin{enumerate}
    \item $\upsilon$\,Sgr is an intermediate-mass star ($M_\mathrm{ZAMS,1}\approx 5\,\mathrm{M}_\odot$) that is not massive enough to explode as a Type Ib supernova, contrary to previous claims \citep{Dudley1990}.
    \item The system is observed at a second mass-transfer stage, in which the primary is losing the remainder of the hydrogen in its envelope, before becoming a carbon-oxygen white dwarf with a helium-dominated atmosphere $\approx 100\,\mathrm{kyr}$ in the future.
    \item The faint companion is a rapidly-rotating B-type star that has gained a significant amount of mass by accretion from the primary.
\end{enumerate}

$\upsilon$\,Sgr is a unique system, containing a rare helium supergiant in the process of being stripped of its remaining envelope, on the way to becoming a white dwarf in the not-too-distant future. Relatively nearby ($d\approx 500\,\mathrm{pc}$), it is a gift of nature that allows us to study in detail key processes of binary evolution, such as envelope stripping of the primary and mass accretion by the companion. Furthering our understanding of $\upsilon$\,Sgr, or finding other systems like it, will help us better understand the evolution of interacting binary stars.

\section*{Acknowledgments}

We thank an anonymous referee for a constructive review of the paper. We thank I.~Arcavi, M.-C.~Lam, N.~Shitrit, and J.~Bodensteiner for helpful comments and discussions. AG acknowledges support from a grant by the Prof.\ Amnon Pazy Research Foundation.  TS acknowledges support from the European Union's Horizon 2020 under the Marie Skłodowska-Curie grant agreement No 101024605. The PoWR code has been developed under the guidance of Wolf-Rainer Hamann with substantial contributions from Lars Koesterke, Gotz Gr\"afener, Andreas Sander, Tomer Shenar and other co-workers and students.

\section*{Data Availability Statement}

The data underlying this article will be made available upon publication.

\bibliographystyle{mnras}
\bibliography{references}

\providecommand{\noopsort}[1]{}
\begin{thebibliography}{}
\makeatletter
\relax
\def\mn@urlcharsother{\let\do\@makeother \do\$\do\&\do\#\do\^\do\_\do\%\do\~}
\def\mn@doi{\begingroup\mn@urlcharsother \@ifnextchar [ {\mn@doi@}
  {\mn@doi@[]}}
\def\mn@doi@[#1]#2{\def\@tempa{#1}\ifx\@tempa\@empty \href
  {http://dx.doi.org/#2} {doi:#2}\else \href {http://dx.doi.org/#2} {#1}\fi
  \endgroup}
\def\mn@eprint#1#2{\mn@eprint@#1:#2::\@nil}
\def\mn@eprint@arXiv#1{\href {http://arxiv.org/abs/#1} {{\tt arXiv:#1}}}
\def\mn@eprint@dblp#1{\href {http://dblp.uni-trier.de/rec/bibtex/#1.xml}
  {dblp:#1}}
\def\mn@eprint@#1:#2:#3:#4\@nil{\def\@tempa {#1}\def\@tempb {#2}\def\@tempc
  {#3}\ifx \@tempc \@empty \let \@tempc \@tempb \let \@tempb \@tempa \fi \ifx
  \@tempb \@empty \def\@tempb {arXiv}\fi \@ifundefined
  {mn@eprint@\@tempb}{\@tempb:\@tempc}{\expandafter \expandafter \csname
  mn@eprint@\@tempb\endcsname \expandafter{\@tempc}}}

\bibitem[\protect\citeauthoryear{{Alastuey} \& {Jancovici}}{{Alastuey} \&
  {Jancovici}}{1978}]{Alastuey1978}
{Alastuey} A.,  {Jancovici} B.,  1978, \mn@doi [\apj] {10.1086/156681}, \href
  {https://ui.adsabs.harvard.edu/\#abs/1978ApJ...226.1034A} {226, 1034}

\bibitem[\protect\citeauthoryear{{Asplund}, {Grevesse}, {Sauval}  \&
  {Scott}}{{Asplund} et~al.}{2009}]{Asplund2009}
{Asplund} M.,  {Grevesse} N.,  {Sauval} A.~J.,   {Scott} P.,  2009, \mn@doi
  [\araa] {10.1146/annurev.astro.46.060407.145222}, \href
  {https://ui.adsabs.harvard.edu/abs/2009ARA&A..47..481A} {47, 481}

\bibitem[\protect\citeauthoryear{{Bailer-Jones}, {Rybizki}, {Fouesneau},
  {Demleitner}  \& {Andrae}}{{Bailer-Jones} et~al.}{2021}]{Bailer-Jones2021}
{Bailer-Jones} C.~A.~L.,  {Rybizki} J.,  {Fouesneau} M.,  {Demleitner} M.,
  {Andrae} R.,  2021, \mn@doi [\aj] {10.3847/1538-3881/abd806}, \href
  {https://ui.adsabs.harvard.edu/abs/2021AJ....161..147B} {161, 147}

\bibitem[\protect\citeauthoryear{{Baum}, {Hamann}, {Koesterke}  \&
  {Wessolowski}}{{Baum} et~al.}{1992}]{Baum1992}
{Baum} E.,  {Hamann} W.~R.,  {Koesterke} L.,   {Wessolowski} U.,  1992, \aap,
  \href {https://ui.adsabs.harvard.edu/abs/1992A&A...266..402B} {266, 402}

\bibitem[\protect\citeauthoryear{{Bodensteiner} et~al.,}{{Bodensteiner}
  et~al.}{2020}]{Bodensteiner2020BeHR}
{Bodensteiner} J.,  et~al., 2020, \mn@doi [\aap] {10.1051/0004-6361/202038682},
  \href {https://ui.adsabs.harvard.edu/abs/2020A&A...641A..43B} {641, A43}

\bibitem[\protect\citeauthoryear{{B{\"o}hm-Vitense}}{{B{\"o}hm-Vitense}}{1958}]{MLT1958}
{B{\"o}hm-Vitense} E.,  1958, \zap, \href
  {https://ui.adsabs.harvard.edu/abs/1958ZA.....46..108B} {46, 108}

\bibitem[\protect\citeauthoryear{{Bonneau} et~al.,}{{Bonneau}
  et~al.}{2011}]{Bonneau2011}
{Bonneau} D.,  et~al., 2011, \mn@doi [\aap] {10.1051/0004-6361/201116742},
  \href {https://ui.adsabs.harvard.edu/abs/2011A&A...532A.148B} {532, A148}

\bibitem[\protect\citeauthoryear{{Buchler} \& {Yueh}}{{Buchler} \&
  {Yueh}}{1976}]{Buchler1976}
{Buchler} J.~R.,  {Yueh} W.~R.,  1976, \mn@doi [\apj] {10.1086/154847}, \href
  {https://ui.adsabs.harvard.edu/abs/1976ApJ...210..440B} {210, 440}

\bibitem[\protect\citeauthoryear{{Campbell}}{{Campbell}}{1895}]{Campbell1895}
{Campbell} W.~W.,  1895, \mn@doi [\apj] {10.1086/140127}, \href
  {https://ui.adsabs.harvard.edu/abs/1895ApJ.....2..177C} {2, 177}

\bibitem[\protect\citeauthoryear{{Campbell}}{{Campbell}}{1899}]{Campbell1899}
{Campbell} W.~W.,  1899, \mn@doi [\apj] {10.1086/140640}, \href
  {https://ui.adsabs.harvard.edu/abs/1899ApJ....10..241C} {10, 241}

\bibitem[\protect\citeauthoryear{{Cannon} \& {Pickering}}{{Cannon} \&
  {Pickering}}{1912}]{Cannon1912}
{Cannon} A.~J.,  {Pickering} E.~C.,  1912, Annals of Harvard College
  Observatory, \href {https://ui.adsabs.harvard.edu/abs/1912AnHar..56...65C}
  {56, 65}

\bibitem[\protect\citeauthoryear{{Cardelli}, {Clayton}  \& {Mathis}}{{Cardelli}
  et~al.}{1989}]{Cardelli1989}
{Cardelli} J.~A.,  {Clayton} G.~C.,   {Mathis} J.~S.,  1989, \mn@doi [\apj]
  {10.1086/167900}, \href
  {https://ui.adsabs.harvard.edu/abs/1989ApJ...345..245C} {345, 245}

\bibitem[\protect\citeauthoryear{{Cassinelli} \& {Olson}}{{Cassinelli} \&
  {Olson}}{1979}]{Cassinelli1979}
{Cassinelli} J.~P.,  {Olson} G.~L.,  1979, \mn@doi [\apj] {10.1086/156956},
  \href {https://ui.adsabs.harvard.edu/abs/1979ApJ...229..304C} {229, 304}

\bibitem[\protect\citeauthoryear{{Cassisi}, {Potekhin}, {Pietrinferni},
  {Catelan}  \& {Salaris}}{{Cassisi} et~al.}{2007}]{Cassisi2007}
{Cassisi} S.,  {Potekhin} A.~Y.,  {Pietrinferni} A.,  {Catelan} M.,   {Salaris}
  M.,  2007, \mn@doi [\apj] {10.1086/516819}, \href
  {https://ui.adsabs.harvard.edu/abs/2007ApJ...661.1094C} {661, 1094}

\bibitem[\protect\citeauthoryear{{Castor}, {Abbott}  \& {Klein}}{{Castor}
  et~al.}{1975}]{CAK1975}
{Castor} J.~I.,  {Abbott} D.~C.,   {Klein} R.~I.,  1975, \mn@doi [\apj]
  {10.1086/153315}, \href
  {https://ui.adsabs.harvard.edu/abs/1975ApJ...195..157C} {195, 157}

\bibitem[\protect\citeauthoryear{{Cyburt} et~al.,}{{Cyburt}
  et~al.}{2010}]{Cyburt2010}
{Cyburt} R.~H.,  et~al., 2010, \mn@doi [\apjs] {10.1088/0067-0049/189/1/240},
  \href {https://ui.adsabs.harvard.edu/abs/2010ApJS..189..240C} {189, 240}

\bibitem[\protect\citeauthoryear{{\noopsort{DeBoer}}{deBoer}
  et~al.,}{{\noopsort{DeBoer}}{deBoer} et~al.}{2017}]{deBoer2017}
{\noopsort{DeBoer}}{deBoer} R.~J.,  et~al., 2017, \mn@doi [Reviews of Modern
  Physics] {10.1103/RevModPhys.89.035007}, \href
  {https://ui.adsabs.harvard.edu/abs/2017RvMP...89c5007D} {89, 035007}

\bibitem[\protect\citeauthoryear{{\noopsort{De~Jager}}{de Jager},
  {Nieuwenhuijzen}  \& {van der Hucht}}{{\noopsort{De~Jager}}{de Jager}
  et~al.}{1988}]{deJager1988}
{\noopsort{De~Jager}}{de Jager} C.,  {Nieuwenhuijzen} H.,   {van der Hucht}
  K.~A.,  1988, \aaps, \href
  {https://ui.adsabs.harvard.edu/abs/1988A&AS...72..259D} {72, 259}

\bibitem[\protect\citeauthoryear{{\noopsort{De~Mink}}{de Mink}, {Sana},
  {Langer}, {Izzard}  \& {Schneider}}{{\noopsort{De~Mink}}{de Mink}
  et~al.}{2014}]{deMink2014}
{\noopsort{De~Mink}}{de Mink} S.~E.,  {Sana} H.,  {Langer} N.,  {Izzard} R.~G.,
    {Schneider} F.~R.~N.,  2014, \mn@doi [\apj] {10.1088/0004-637X/782/1/7},
  \href {https://ui.adsabs.harvard.edu/abs/2014ApJ...782....7D} {782, 7}

\bibitem[\protect\citeauthoryear{{Dewitt}, {Graboske}  \& {Cooper}}{{Dewitt}
  et~al.}{1973}]{Dewitt1973}
{Dewitt} H.~E.,  {Graboske} H.~C.,   {Cooper} M.~S.,  1973, \mn@doi [\apj]
  {10.1086/152061}, \href
  {https://ui.adsabs.harvard.edu/\#abs/1973ApJ...181..439D} {181, 439}

\bibitem[\protect\citeauthoryear{{Dionne} \& {Robert}}{{Dionne} \&
  {Robert}}{2006}]{Dionne2006}
{Dionne} D.,  {Robert} C.,  2006, \mn@doi [\apj] {10.1086/500380}, \href
  {https://ui.adsabs.harvard.edu/abs/2006ApJ...641..252D} {641, 252}

\bibitem[\protect\citeauthoryear{{Drilling} \& {Schonberner}}{{Drilling} \&
  {Schonberner}}{1982}]{Drilling1982}
{Drilling} J.~S.,  {Schonberner} D.,  1982, \aap, \href
  {https://ui.adsabs.harvard.edu/abs/1982A&A...113L..22D} {113, L22}

\bibitem[\protect\citeauthoryear{{Dsilva}, {Shenar}, {Sana}  \&
  {Marchant}}{{Dsilva} et~al.}{2020}]{Dsilva2020}
{Dsilva} K.,  {Shenar} T.,  {Sana} H.,   {Marchant} P.,  2020, \mn@doi [\aap]
  {10.1051/0004-6361/202038446}, \href
  {https://ui.adsabs.harvard.edu/abs/2020A&A...641A..26D} {641, A26}

\bibitem[\protect\citeauthoryear{{Dudley}}{{Dudley}}{1992}]{Dudley1992}
{Dudley} R.~E.,  1992, PhD thesis, Saint Andrews University, UK

\bibitem[\protect\citeauthoryear{{Dudley} \& {Jeffery}}{{Dudley} \&
  {Jeffery}}{1990}]{Dudley1990}
{Dudley} R.~E.,  {Jeffery} C.~S.,  1990, \mnras, \href
  {https://ui.adsabs.harvard.edu/abs/1990MNRAS.247..400D} {247, 400}

\bibitem[\protect\citeauthoryear{{Dudley} \& {Jeffery}}{{Dudley} \&
  {Jeffery}}{1993}]{Dudley1993}
{Dudley} R.~E.,  {Jeffery} C.~S.,  1993, \mn@doi [\mnras]
  {10.1093/mnras/262.4.945}, \href
  {https://ui.adsabs.harvard.edu/abs/1993MNRAS.262..945D} {262, 945}

\bibitem[\protect\citeauthoryear{{Duvignau}, {Friedjung}  \& {Hack}}{{Duvignau}
  et~al.}{1979}]{Duvignau1979}
{Duvignau} H.,  {Friedjung} M.,   {Hack} M.,  1979, \aap, \href
  {https://ui.adsabs.harvard.edu/abs/1979A&A....71..310D} {71, 310}

\bibitem[\protect\citeauthoryear{{Eggleton}}{{Eggleton}}{1983}]{Eggleton1983}
{Eggleton} P.~P.,  1983, \mn@doi [\apj] {10.1086/160960}, \href
  {https://ui.adsabs.harvard.edu/abs/1983ApJ...268..368E} {268, 368}

\bibitem[\protect\citeauthoryear{{El-Badry} \& {Burdge}}{{El-Badry} \&
  {Burdge}}{2022}]{El-Badry2022}
{El-Badry} K.,  {Burdge} K.~B.,  2022, \mn@doi [\mnras]
  {10.1093/mnrasl/slab135}, \href
  {https://ui.adsabs.harvard.edu/abs/2022MNRAS.511L..24E} {511, 24}

\bibitem[\protect\citeauthoryear{{Feldmeier}, {Puls}  \&
  {Pauldrach}}{{Feldmeier} et~al.}{1997}]{Feldmeier1997}
{Feldmeier} A.,  {Puls} J.,   {Pauldrach} A.~W.~A.,  1997, \aap, \href
  {https://ui.adsabs.harvard.edu/abs/1997A&A...322..878F} {322, 878}

\bibitem[\protect\citeauthoryear{{Ferguson}, {Alexander}, {Allard}, {Barman},
  {Bodnarik}, {Hauschildt}, {Heffner-Wong}  \& {Tamanai}}{{Ferguson}
  et~al.}{2005}]{Ferguson2005}
{Ferguson} J.~W.,  {Alexander} D.~R.,  {Allard} F.,  {Barman} T.,  {Bodnarik}
  J.~G.,  {Hauschildt} P.~H.,  {Heffner-Wong} A.,   {Tamanai} A.,  2005,
  \mn@doi [\apj] {10.1086/428642}, \href
  {https://ui.adsabs.harvard.edu/abs/2005ApJ...623..585F} {623, 585}

\bibitem[\protect\citeauthoryear{{Friend} \& {Abbott}}{{Friend} \&
  {Abbott}}{1986}]{Friend1986}
{Friend} D.~B.,  {Abbott} D.~C.,  1986, \mn@doi [\apj] {10.1086/164809}, \href
  {https://ui.adsabs.harvard.edu/abs/1986ApJ...311..701F} {311, 701}

\bibitem[\protect\citeauthoryear{{Fuller}, {Fowler}  \& {Newman}}{{Fuller}
  et~al.}{1985}]{Fuller1985}
{Fuller} G.~M.,  {Fowler} W.~A.,   {Newman} M.~J.,  1985, \mn@doi [\apj]
  {10.1086/163208}, \href
  {https://ui.adsabs.harvard.edu/abs/1985ApJ...293....1F} {293, 1}

\bibitem[\protect\citeauthoryear{{Gaia Collaboration} et~al.,}{{Gaia
  Collaboration} et~al.}{2022}]{GaiaDR3}
{Gaia Collaboration} et~al., 2022, arXiv e-prints, \href
  {https://ui.adsabs.harvard.edu/abs/2022arXiv220800211G} {p. arXiv:2208.00211}

\bibitem[\protect\citeauthoryear{{Gal-Yam} et~al.,}{{Gal-Yam}
  et~al.}{2022}]{Gal-Yam2022}
{Gal-Yam} A.,  et~al., 2022, \mn@doi [\nat] {10.1038/s41586-021-04155-1}, \href
  {https://ui.adsabs.harvard.edu/abs/2022Natur.601..201G} {601, 201}

\bibitem[\protect\citeauthoryear{{Gilkis}, {Vink}, {Eldridge}  \&
  {Tout}}{{Gilkis} et~al.}{2019}]{GilkisVinkEldridgeTout2019}
{Gilkis} A.,  {Vink} J.~S.,  {Eldridge} J.~J.,   {Tout} C.~A.,  2019, \mn@doi
  [\mnras] {10.1093/mnras/stz1134}, \href
  {https://ui.adsabs.harvard.edu/abs/2019MNRAS.486.4451G} {486, 4451}

\bibitem[\protect\citeauthoryear{{G{\"o}tberg}, {de Mink}  \&
  {Groh}}{{G{\"o}tberg} et~al.}{2017}]{Goetberg2017}
{G{\"o}tberg} Y.,  {de Mink} S.~E.,   {Groh} J.~H.,  2017, \mn@doi [\aap]
  {10.1051/0004-6361/201730472}, \href
  {https://ui.adsabs.harvard.edu/abs/2017A&A...608A..11G} {608, A11}

\bibitem[\protect\citeauthoryear{{Gr{\"a}fener}, {Koesterke}  \&
  {Hamann}}{{Gr{\"a}fener} et~al.}{2002}]{PoWR2002}
{Gr{\"a}fener} G.,  {Koesterke} L.,   {Hamann} W.~R.,  2002, \mn@doi [\aap]
  {10.1051/0004-6361:20020269}, \href
  {https://ui.adsabs.harvard.edu/abs/2002A&A...387..244G} {387, 244}

\bibitem[\protect\citeauthoryear{{Gray}}{{Gray}}{1992}]{Gray1992}
{Gray} D.~F.,  1992, {The observation and analysis of stellar photospheres.}.
 Vol. 20, Cambridge University Press

\bibitem[\protect\citeauthoryear{{Greenstein}}{{Greenstein}}{1940}]{Greenstein1940}
{Greenstein} J.~L.,  1940, \mn@doi [\apj] {10.1086/144185}, \href
  {https://ui.adsabs.harvard.edu/abs/1940ApJ....91..438G} {91, 438}

\bibitem[\protect\citeauthoryear{{Groh}, {Oliveira}  \& {Steiner}}{{Groh}
  et~al.}{2008}]{Groh2008}
{Groh} J.~H.,  {Oliveira} A.~S.,   {Steiner} J.~E.,  2008, \mn@doi [\aap]
  {10.1051/0004-6361:200809511}, \href
  {https://ui.adsabs.harvard.edu/abs/2008A&A...485..245G} {485, 245}

\bibitem[\protect\citeauthoryear{{Hack}}{{Hack}}{1966}]{Hack1966}
{Hack} M.,  1966, in {Hubenet} H.,  ed.,  Vol. 26, Abundance Determinations in
  Stellar Spectra. p.~227

\bibitem[\protect\citeauthoryear{{Hack}}{{Hack}}{1980}]{Hack1980}
{Hack} M.,  1980, in {Plavec} M.~J.,  {Popper} D.~M.,   {Ulrich} R.~K.,  eds,
  Vol. 88, Close Binary Stars: Observations and Interpretation. pp 271--285

\bibitem[\protect\citeauthoryear{{Hack} \& {Pasinetti}}{{Hack} \&
  {Pasinetti}}{1963}]{Hack1963}
{Hack} M.,  {Pasinetti} L.,  1963, Contr. Oss. Astr. Milano-Merate, 215, 1

\bibitem[\protect\citeauthoryear{{Hadrava}}{{Hadrava}}{1995}]{Hadrava1995}
{Hadrava} P.,  1995, \aaps, \href
  {https://ui.adsabs.harvard.edu/abs/1995A&AS..114..393H} {114, 393}

\bibitem[\protect\citeauthoryear{{Hainich}, {Ramachandran}, {Shenar}, {Sander},
  {Todt}, {Gruner}, {Oskinova}  \& {Hamann}}{{Hainich}
  et~al.}{2019}]{Hainich2019}
{Hainich} R.,  {Ramachandran} V.,  {Shenar} T.,  {Sander} A.~A.~C.,  {Todt} H.,
   {Gruner} D.,  {Oskinova} L.~M.,   {Hamann} W.~R.,  2019, \mn@doi [\aap]
  {10.1051/0004-6361/201833787}, \href
  {https://ui.adsabs.harvard.edu/abs/2019A&A...621A..85H} {621, A85}

\bibitem[\protect\citeauthoryear{{Halabi}, {Izzard}  \& {Tout}}{{Halabi}
  et~al.}{2018}]{Halabi2018}
{Halabi} G.~M.,  {Izzard} R.~G.,   {Tout} C.~A.,  2018, \mn@doi [\mnras]
  {10.1093/mnras/sty2243}, \href
  {https://ui.adsabs.harvard.edu/abs/2018MNRAS.480.5176H} {480, 5176}

\bibitem[\protect\citeauthoryear{{Hamann} \& {Gr{\"a}fener}}{{Hamann} \&
  {Gr{\"a}fener}}{2003}]{PoWR2003}
{Hamann} W.~R.,  {Gr{\"a}fener} G.,  2003, \mn@doi [\aap]
  {10.1051/0004-6361:20031308}, \href
  {https://ui.adsabs.harvard.edu/abs/2003A&A...410..993H} {410, 993}

\bibitem[\protect\citeauthoryear{{Hamann} et~al.,}{{Hamann}
  et~al.}{2019}]{Hamann2019}
{Hamann} W.~R.,  et~al., 2019, \mn@doi [\aap] {10.1051/0004-6361/201834850},
  \href {https://ui.adsabs.harvard.edu/abs/2019A&A...625A..57H} {625, A57}

\bibitem[\protect\citeauthoryear{{Henyey}, {Vardya}  \& {Bodenheimer}}{{Henyey}
  et~al.}{1965}]{MLT}
{Henyey} L.,  {Vardya} M.~S.,   {Bodenheimer} P.,  1965, \mn@doi [\apj]
  {10.1086/148357}, \href
  {https://ui.adsabs.harvard.edu/abs/1965ApJ...142..841H} {142, 841}

\bibitem[\protect\citeauthoryear{{Herwig}}{{Herwig}}{2000}]{Herwig2000}
{Herwig} F.,  2000, \aap, \href
  {https://ui.adsabs.harvard.edu/abs/2000A&A...360..952H} {360, 952}

\bibitem[\protect\citeauthoryear{{Hurley}, {Pols}  \& {Tout}}{{Hurley}
  et~al.}{2000}]{Hurley2000}
{Hurley} J.~R.,  {Pols} O.~R.,   {Tout} C.~A.,  2000, \mn@doi [\mnras]
  {10.1046/j.1365-8711.2000.03426.x}, \href
  {https://ui.adsabs.harvard.edu/abs/2000MNRAS.315..543H} {315, 543}

\bibitem[\protect\citeauthoryear{{Hutter}, {Tycner}, {Zavala}, {Benson},
  {Hummel}  \& {Zirm}}{{Hutter} et~al.}{2021}]{Hutter2021}
{Hutter} D.~J.,  {Tycner} C.,  {Zavala} R.~T.,  {Benson} J.~A.,  {Hummel}
  C.~A.,   {Zirm} H.,  2021, \mn@doi [\apjs] {10.3847/1538-4365/ac23cb}, \href
  {https://ui.adsabs.harvard.edu/abs/2021ApJS..257...69H} {257, 69}

\bibitem[\protect\citeauthoryear{{Iglesias} \& {Rogers}}{{Iglesias} \&
  {Rogers}}{1993}]{Iglesias1993}
{Iglesias} C.~A.,  {Rogers} F.~J.,  1993, \mn@doi [\apj] {10.1086/172958},
  \href {https://ui.adsabs.harvard.edu/abs/1993ApJ...412..752I} {412, 752}

\bibitem[\protect\citeauthoryear{{Iglesias} \& {Rogers}}{{Iglesias} \&
  {Rogers}}{1996}]{Iglesias1996}
{Iglesias} C.~A.,  {Rogers} F.~J.,  1996, \mn@doi [\apj] {10.1086/177381},
  \href {https://ui.adsabs.harvard.edu/abs/1996ApJ...464..943I} {464, 943}

\bibitem[\protect\citeauthoryear{{Irwin}}{{Irwin}}{2004}]{Irwin2004}
{Irwin} A.~W.,  2004, The FreeEOS Code for Calculating the Equation of State
  for Stellar Interiors, \url {http://freeeos.sourceforge.net/}

\bibitem[\protect\citeauthoryear{{Itoh}, {Totsuji}, {Ichimaru}  \&
  {Dewitt}}{{Itoh} et~al.}{1979}]{Itoh1979}
{Itoh} N.,  {Totsuji} H.,  {Ichimaru} S.,   {Dewitt} H.~E.,  1979, \mn@doi
  [\apj] {10.1086/157590}, \href
  {https://ui.adsabs.harvard.edu/\#abs/1979ApJ...234.1079I} {234, 1079}

\bibitem[\protect\citeauthoryear{{Itoh}, {Hayashi}, {Nishikawa}  \&
  {Kohyama}}{{Itoh} et~al.}{1996}]{Itoh1996}
{Itoh} N.,  {Hayashi} H.,  {Nishikawa} A.,   {Kohyama} Y.,  1996, \mn@doi
  [\apjs] {10.1086/192264}, \href
  {https://ui.adsabs.harvard.edu/abs/1996ApJS..102..411I} {102, 411}

\bibitem[\protect\citeauthoryear{{Jeffery}}{{Jeffery}}{1988}]{Jeffery1988}
{Jeffery} C.~S.,  1988, \mn@doi [\mnras] {10.1093/mnras/235.4.1287}, \href
  {https://ui.adsabs.harvard.edu/abs/1988MNRAS.235.1287J} {235, 1287}

\bibitem[\protect\citeauthoryear{{Kolb} \& {Ritter}}{{Kolb} \&
  {Ritter}}{1990}]{Kolb1990}
{Kolb} U.,  {Ritter} H.,  1990, \aap, \href
  {https://ui.adsabs.harvard.edu/abs/1990A&A...236..385K} {236, 385}

\bibitem[\protect\citeauthoryear{{Koubsk{\'y}}, {Harmanec}, {Yang},
  {Netolick{\'y}}, {{\v{S}}koda}, {{\v{S}}lechta}  \&
  {Kor{\v{c}}{\'a}kov{\'a}}}{{Koubsk{\'y}} et~al.}{2006}]{Koubsky2006}
{Koubsk{\'y}} P.,  {Harmanec} P.,  {Yang} S.,  {Netolick{\'y}} M.,
  {{\v{S}}koda} P.,  {{\v{S}}lechta} M.,   {Kor{\v{c}}{\'a}kov{\'a}} D.,  2006,
  \mn@doi [\aap] {10.1051/0004-6361:20065274}, \href
  {https://ui.adsabs.harvard.edu/abs/2006A&A...459..849K} {459, 849}

\bibitem[\protect\citeauthoryear{{Koubsk{\'y}}, {Harmanec}, {Yang},
  {Kor{\v{c}}{\'a}kov{\'a}}, {Netolick{\'y}}, {{\v{S}}koda}, {{\v{S}}lechta}
  \& {Votruba}}{{Koubsk{\'y}} et~al.}{2007}]{Koubsky2007}
{Koubsk{\'y}} P.,  {Harmanec} P.,  {Yang} S.,  {Kor{\v{c}}{\'a}kov{\'a}} D.,
  {Netolick{\'y}} M.,  {{\v{S}}koda} P.,  {{\v{S}}lechta} M.,   {Votruba} V.,
  2007, in {Demircan} O.,  {Selam} S.~O.,   {Albayrak} B.,  eds,  Astronomical
  Society of the Pacific Conference Series Vol. 370, Solar and Stellar Physics
  Through Eclipses. p.~207

\bibitem[\protect\citeauthoryear{{Langanke} \&
  {Mart{\'{\i}}nez-Pinedo}}{{Langanke} \&
  {Mart{\'{\i}}nez-Pinedo}}{2000}]{Langanke2000}
{Langanke} K.,  {Mart{\'{\i}}nez-Pinedo} G.,  2000, \mn@doi [Nuclear Physics A]
  {10.1016/S0375-9474(00)00131-7}, \href
  {https://ui.adsabs.harvard.edu/abs/2000NuPhA.673..481L} {673, 481}

\bibitem[\protect\citeauthoryear{{Laplace}, {G{\"o}tberg}, {de Mink}, {Justham}
   \& {Farmer}}{{Laplace} et~al.}{2020}]{Laplace2020}
{Laplace} E.,  {G{\"o}tberg} Y.,  {de Mink} S.~E.,  {Justham} S.,   {Farmer}
  R.,  2020, \mn@doi [\aap] {10.1051/0004-6361/201937300}, \href
  {https://ui.adsabs.harvard.edu/abs/2020A&A...637A...6L} {637, A6}

\bibitem[\protect\citeauthoryear{{Leushin}}{{Leushin}}{2000}]{Leushin2000}
{Leushin} V.~V.,  2000, Bulletin of the Special Astrophysics Observatory, \href
  {https://ui.adsabs.harvard.edu/abs/2000BSAO...50...60L} {50, 60}

\bibitem[\protect\citeauthoryear{{Mahy} et~al.,}{{Mahy}
  et~al.}{2022}]{Mahy2022}
{Mahy} L.,  et~al., 2022, \mn@doi [\aap] {10.1051/0004-6361/202243147}, \href
  {https://ui.adsabs.harvard.edu/abs/2022A&A...664A.159M} {664, A159}

\bibitem[\protect\citeauthoryear{{Massey}}{{Massey}}{1981}]{Massey1981}
{Massey} P.,  1981, \mn@doi [\apj] {10.1086/158908}, \href
  {https://ui.adsabs.harvard.edu/abs/1981ApJ...246..153M} {246, 153}

\bibitem[\protect\citeauthoryear{{Mehta}, {Buonanno}, {Gair}, {Miller},
  {Farag}, {deBoer}, {Wiescher}  \& {Timmes}}{{Mehta} et~al.}{2022}]{Mehta2022}
{Mehta} A.~K.,  {Buonanno} A.,  {Gair} J.,  {Miller} M.~C.,  {Farag} E.,
  {deBoer} R.~J.,  {Wiescher} M.,   {Timmes} F.~X.,  2022, \mn@doi [\apj]
  {10.3847/1538-4357/ac3130}, \href
  {https://ui.adsabs.harvard.edu/abs/2022ApJ...924...39M} {924, 39}

\bibitem[\protect\citeauthoryear{{Mirabel} \& {Rodrigues}}{{Mirabel} \&
  {Rodrigues}}{2003}]{Mirabel2003}
{Mirabel} I.~F.,  {Rodrigues} I.,  2003, \mn@doi [Science]
  {10.1126/science.1083451}, \href
  {https://ui.adsabs.harvard.edu/abs/2003Sci...300.1119M} {300, 1119}

\bibitem[\protect\citeauthoryear{{Morel} \& {Magnenat}}{{Morel} \&
  {Magnenat}}{1978}]{Morel1978}
{Morel} M.,  {Magnenat} P.,  1978, \aaps, \href
  {https://ui.adsabs.harvard.edu/abs/1978A&AS...34..477M} {34, 477}

\bibitem[\protect\citeauthoryear{{Netolick{\'y}}, {Bonneau}, {Chesneau},
  {Harmanec}, {Koubsk{\'y}}, {Mourard}  \& {Stee}}{{Netolick{\'y}}
  et~al.}{2009}]{Netolicky2009}
{Netolick{\'y}} M.,  {Bonneau} D.,  {Chesneau} O.,  {Harmanec} P.,
  {Koubsk{\'y}} P.,  {Mourard} D.,   {Stee} P.,  2009, \mn@doi [\aap]
  {10.1051/0004-6361/200811192}, \href
  {https://ui.adsabs.harvard.edu/abs/2009A&A...499..827N} {499, 827}

\bibitem[\protect\citeauthoryear{{Newville}, {Stensitzki}, {Allen}  \&
  {Ingargiola}}{{Newville} et~al.}{2014}]{Newville2014}
{Newville} M.,  {Stensitzki} T.,  {Allen} D.~B.,   {Ingargiola} A.,  2014,
  {LMFIT: Non-Linear Least-Square Minimization and Curve-Fitting for Python},
  Zenodo, \mn@doi{10.5281/zenodo.11813}

\bibitem[\protect\citeauthoryear{{Oda}, {Hino}, {Muto}, {Takahara}  \&
  {Sato}}{{Oda} et~al.}{1994}]{Oda1994}
{Oda} T.,  {Hino} M.,  {Muto} K.,  {Takahara} M.,   {Sato} K.,  1994, \mn@doi
  [Atomic Data and Nuclear Data Tables] {10.1006/adnd.1994.1007}, \href
  {https://ui.adsabs.harvard.edu/abs/1994ADNDT..56..231O} {56, 231}

\bibitem[\protect\citeauthoryear{{Oskinova}, {Feldmeier}  \&
  {Hamann}}{{Oskinova} et~al.}{2006}]{Oskinova2006}
{Oskinova} L.~M.,  {Feldmeier} A.,   {Hamann} W.~R.,  2006, \mn@doi [\mnras]
  {10.1111/j.1365-2966.2006.10858.x}, \href
  {https://ui.adsabs.harvard.edu/abs/2006MNRAS.372..313O} {372, 313}

\bibitem[\protect\citeauthoryear{{Packet}}{{Packet}}{1981}]{Packet1981}
{Packet} W.,  1981, \aap, \href
  {https://ui.adsabs.harvard.edu/abs/1981A&A...102...17P} {102, 17}

\bibitem[\protect\citeauthoryear{{Paczy{\'n}ski}}{{Paczy{\'n}ski}}{1967}]{Paczynski1967}
{Paczy{\'n}ski} B.,  1967, \actaa, \href
  {https://ui.adsabs.harvard.edu/abs/1967AcA....17..355P} {17, 355}

\bibitem[\protect\citeauthoryear{{Paczy{\'n}ski}}{{Paczy{\'n}ski}}{1971a}]{Paczynski1971a}
{Paczy{\'n}ski} B.,  1971a, \mn@doi [\araa]
  {10.1146/annurev.aa.09.090171.001151}, \href
  {https://ui.adsabs.harvard.edu/abs/1971ARA&A...9..183P} {9, 183}

\bibitem[\protect\citeauthoryear{{Paczy{\'n}ski}}{{Paczy{\'n}ski}}{1971b}]{Paczynski1971b}
{Paczy{\'n}ski} B.,  1971b, \actaa, \href
  {https://ui.adsabs.harvard.edu/abs/1971AcA....21....1P} {21, 1}

\bibitem[\protect\citeauthoryear{{Parthasarathy}, {Cornachin}  \&
  {Hack}}{{Parthasarathy} et~al.}{1986}]{Parthasarathy1986}
{Parthasarathy} M.,  {Cornachin} M.,   {Hack} M.,  1986, \aap, \href
  {https://ui.adsabs.harvard.edu/abs/1986A&A...166..237P} {166, 237}

\bibitem[\protect\citeauthoryear{{Parthasarathy}, {Hack}  \&
  {Tektunali}}{{Parthasarathy} et~al.}{1990}]{Parthasarathy1990}
{Parthasarathy} M.,  {Hack} M.,   {Tektunali} G.,  1990, \aap, \href
  {https://ui.adsabs.harvard.edu/abs/1990A&A...230..136P} {230, 136}

\bibitem[\protect\citeauthoryear{{Paxton}, {Bildsten}, {Dotter}, {Herwig},
  {Lesaffre}  \& {Timmes}}{{Paxton} et~al.}{2011}]{Paxton2011}
{Paxton} B.,  {Bildsten} L.,  {Dotter} A.,  {Herwig} F.,  {Lesaffre} P.,
  {Timmes} F.,  2011, \mn@doi [\apjs] {10.1088/0067-0049/192/1/3}, \href
  {https://ui.adsabs.harvard.edu/abs/2011ApJS..192....3P} {192, 3}

\bibitem[\protect\citeauthoryear{{Paxton} et~al.,}{{Paxton}
  et~al.}{2013}]{Paxton2013}
{Paxton} B.,  et~al., 2013, \mn@doi [\apjs] {10.1088/0067-0049/208/1/4}, \href
  {https://ui.adsabs.harvard.edu/abs/2013ApJS..208....4P} {208, 4}

\bibitem[\protect\citeauthoryear{{Paxton} et~al.,}{{Paxton}
  et~al.}{2015}]{Paxton2015}
{Paxton} B.,  et~al., 2015, \mn@doi [\apjs] {10.1088/0067-0049/220/1/15}, \href
  {https://ui.adsabs.harvard.edu/abs/2015ApJS..220...15P} {220, 15}

\bibitem[\protect\citeauthoryear{{Paxton} et~al.,}{{Paxton}
  et~al.}{2018}]{Paxton2018}
{Paxton} B.,  et~al., 2018, \mn@doi [\apjs] {10.3847/1538-4365/aaa5a8}, \href
  {https://ui.adsabs.harvard.edu/abs/2018ApJS..234...34P} {234, 34}

\bibitem[\protect\citeauthoryear{{Paxton} et~al.,}{{Paxton}
  et~al.}{2019}]{Paxton2019}
{Paxton} B.,  et~al., 2019, \mn@doi [\apjs] {10.3847/1538-4365/ab2241}, \href
  {https://ui.adsabs.harvard.edu/abs/2019ApJS..243...10P} {243, 10}

\bibitem[\protect\citeauthoryear{{Podsiadlowski}, {Joss}  \&
  {Hsu}}{{Podsiadlowski} et~al.}{1992}]{Podsiadlowski1992}
{Podsiadlowski} P.,  {Joss} P.~C.,   {Hsu} J.~J.~L.,  1992, \mn@doi [\apj]
  {10.1086/171341}, \href
  {https://ui.adsabs.harvard.edu/abs/1992ApJ...391..246P} {391, 246}

\bibitem[\protect\citeauthoryear{{Potekhin} \& {Chabrier}}{{Potekhin} \&
  {Chabrier}}{2010}]{Potekhin2010}
{Potekhin} A.~Y.,  {Chabrier} G.,  2010, \mn@doi [Contributions to Plasma
  Physics] {10.1002/ctpp.201010017}, \href
  {https://ui.adsabs.harvard.edu/abs/2010CoPP...50...82P} {50, 82}

\bibitem[\protect\citeauthoryear{{Raskin} et~al.,}{{Raskin}
  et~al.}{2011}]{Raskin2011}
{Raskin} G.,  et~al., 2011, \mn@doi [\aap] {10.1051/0004-6361/201015435}, \href
  {https://ui.adsabs.harvard.edu/abs/2011A&A...526A..69R} {526, A69}

\bibitem[\protect\citeauthoryear{{Renzo} \& {G{\"o}tberg}}{{Renzo} \&
  {G{\"o}tberg}}{2021}]{Renzo2021}
{Renzo} M.,  {G{\"o}tberg} Y.,  2021, \mn@doi [\apj]
  {10.3847/1538-4357/ac29c5}, \href
  {https://ui.adsabs.harvard.edu/abs/2021ApJ...923..277R} {923, 277}

\bibitem[\protect\citeauthoryear{{Rogers} \& {Nayfonov}}{{Rogers} \&
  {Nayfonov}}{2002}]{Rogers2002}
{Rogers} F.~J.,  {Nayfonov} A.,  2002, \mn@doi [\apj] {10.1086/341894}, \href
  {https://ui.adsabs.harvard.edu/abs/2002ApJ...576.1064R} {576, 1064}

\bibitem[\protect\citeauthoryear{{Salpeter}}{{Salpeter}}{1954}]{Salpeter1954}
{Salpeter} E.~E.,  1954, \mn@doi [Australian Journal of Physics]
  {10.1071/PH540373}, \href
  {https://ui.adsabs.harvard.edu/\#abs/1954AuJPh...7..373S} {7, 373}

\bibitem[\protect\citeauthoryear{{Sana} et~al.,}{{Sana}
  et~al.}{2012}]{Sana2012}
{Sana} H.,  et~al., 2012, \mn@doi [Science] {10.1126/science.1223344}, \href
  {https://ui.adsabs.harvard.edu/abs/2012Sci...337..444S} {337, 444}

\bibitem[\protect\citeauthoryear{{Sana} et~al.,}{{Sana}
  et~al.}{2013}]{Sana2013}
{Sana} H.,  et~al., 2013, \mn@doi [\aap] {10.1051/0004-6361/201219621}, \href
  {https://ui.adsabs.harvard.edu/abs/2013A&A...550A.107S} {550, A107}

\bibitem[\protect\citeauthoryear{{Sander}, {Shenar}, {Hainich},
  {G{\'\i}menez-Garc{\'\i}a}, {Todt}  \& {Hamann}}{{Sander}
  et~al.}{2015}]{PoWR2015}
{Sander} A.,  {Shenar} T.,  {Hainich} R.,  {G{\'\i}menez-Garc{\'\i}a} A.,
  {Todt} H.,   {Hamann} W.~R.,  2015, \mn@doi [\aap]
  {10.1051/0004-6361/201425356}, \href
  {https://ui.adsabs.harvard.edu/abs/2015A&A...577A..13S} {577, A13}

\bibitem[\protect\citeauthoryear{{Saumon}, {Chabrier}  \& {van Horn}}{{Saumon}
  et~al.}{1995}]{Saumon1995}
{Saumon} D.,  {Chabrier} G.,   {van Horn} H.~M.,  1995, \mn@doi [\apjs]
  {10.1086/192204}, \href
  {https://ui.adsabs.harvard.edu/abs/1995ApJS...99..713S} {99, 713}

\bibitem[\protect\citeauthoryear{{Schneider}, {Langer}, {de Koter}, {Brott},
  {Izzard}  \& {Lau}}{{Schneider} et~al.}{2014}]{Bonnsai2014}
{Schneider} F.~R.~N.,  {Langer} N.,  {de Koter} A.,  {Brott} I.,  {Izzard}
  R.~G.,   {Lau} H.~H.~B.,  2014, \mn@doi [\aap] {10.1051/0004-6361/201424286},
  \href {https://ui.adsabs.harvard.edu/abs/2014A&A...570A..66S} {570, A66}

\bibitem[\protect\citeauthoryear{{Schoenberner} \& {Drilling}}{{Schoenberner}
  \& {Drilling}}{1983}]{Schoenberner1983}
{Schoenberner} D.,  {Drilling} J.~S.,  1983, \mn@doi [\apj] {10.1086/160947},
  \href {https://ui.adsabs.harvard.edu/abs/1983ApJ...268..225S} {268, 225}

\bibitem[\protect\citeauthoryear{{Schoenberner} \& {Drilling}}{{Schoenberner}
  \& {Drilling}}{1984}]{Schoenberner1984}
{Schoenberner} D.,  {Drilling} J.~S.,  1984, \mn@doi [\apj] {10.1086/161606},
  \href {https://ui.adsabs.harvard.edu/abs/1984ApJ...276..229S} {276, 229}

\bibitem[\protect\citeauthoryear{{Shenar} et~al.,}{{Shenar}
  et~al.}{2015}]{Shenar2015}
{Shenar} T.,  et~al., 2015, \mn@doi [\apj] {10.1088/0004-637X/809/2/135}, \href
  {https://ui.adsabs.harvard.edu/abs/2015ApJ...809..135S} {809, 135}

\bibitem[\protect\citeauthoryear{{Shenar} et~al.,}{{Shenar}
  et~al.}{2017}]{Shenar2017b}
{Shenar} T.,  et~al., 2017, \mn@doi [\aap] {10.1051/0004-6361/201629621}, \href
  {https://ui.adsabs.harvard.edu/abs/2017A&A...598A..85S} {598, A85}

\bibitem[\protect\citeauthoryear{{Shenar}, {Gilkis}, {Vink}, {Sana}  \&
  {Sander}}{{Shenar} et~al.}{2020a}]{Shenar2020WR}
{Shenar} T.,  {Gilkis} A.,  {Vink} J.~S.,  {Sana} H.,   {Sander} A.~A.~C.,
  2020a, \mn@doi [\aap] {10.1051/0004-6361/201936948}, \href
  {https://ui.adsabs.harvard.edu/abs/2020A&A...634A..79S} {634, A79}

\bibitem[\protect\citeauthoryear{{Shenar} et~al.,}{{Shenar}
  et~al.}{2020b}]{Shenar2020}
{Shenar} T.,  et~al., 2020b, \mn@doi [\aap] {10.1051/0004-6361/202038275},
  \href {https://ui.adsabs.harvard.edu/abs/2020A&A...639L...6S} {639, L6}

\bibitem[\protect\citeauthoryear{{Shenar} et~al.,}{{Shenar}
  et~al.}{2022}]{Shenar2022}
{Shenar} T.,  et~al., 2022, \mn@doi [Nature Astronomy]
  {10.1038/s41550-022-01730-y}, \href
  {https://ui.adsabs.harvard.edu/abs/2022NatAs...6.1085S} {6, 1085}

\bibitem[\protect\citeauthoryear{{Sim{\'o}n-D{\'\i}az} \&
  {Herrero}}{{Sim{\'o}n-D{\'\i}az} \& {Herrero}}{2007}]{Simon-Diaz2007}
{Sim{\'o}n-D{\'\i}az} S.,  {Herrero} A.,  2007, \mn@doi [\aap]
  {10.1051/0004-6361:20066060}, \href
  {https://ui.adsabs.harvard.edu/abs/2007A&A...468.1063S} {468, 1063}

\bibitem[\protect\citeauthoryear{{Smartt}}{{Smartt}}{2009}]{Smartt2009}
{Smartt} S.~J.,  2009, \mn@doi [\araa] {10.1146/annurev-astro-082708-101737},
  \href {https://ui.adsabs.harvard.edu/abs/2009ARA&A..47...63S} {47, 63}

\bibitem[\protect\citeauthoryear{{Sukhbold}, {Ertl}, {Woosley}, {Brown}  \&
  {Janka}}{{Sukhbold} et~al.}{2016}]{Sukhbold2016}
{Sukhbold} T.,  {Ertl} T.,  {Woosley} S.~E.,  {Brown} J.~M.,   {Janka} H.~T.,
  2016, \mn@doi [\apj] {10.3847/0004-637X/821/1/38}, \href
  {https://ui.adsabs.harvard.edu/abs/2016ApJ...821...38S} {821, 38}

\bibitem[\protect\citeauthoryear{{Timmes} \& {Swesty}}{{Timmes} \&
  {Swesty}}{2000}]{Timmes2000}
{Timmes} F.~X.,  {Swesty} F.~D.,  2000, \mn@doi [\apjs] {10.1086/313304}, \href
  {https://ui.adsabs.harvard.edu/abs/2000ApJS..126..501T} {126, 501}

\bibitem[\protect\citeauthoryear{{Vink}}{{Vink}}{2017}]{Vink2017}
{Vink} J.~S.,  2017, \mn@doi [\aap] {10.1051/0004-6361/201731902}, \href
  {https://ui.adsabs.harvard.edu/abs/2017A&A...607L...8V} {607, L8}

\bibitem[\protect\citeauthoryear{{Vink} \& {Sander}}{{Vink} \&
  {Sander}}{2021}]{VS21}
{Vink} J.~S.,  {Sander} A. A.~C.,  2021, \mn@doi [\mnras]
  {10.1093/mnras/stab902}, \href
  {https://ui.adsabs.harvard.edu/abs/2021MNRAS.504.2051V} {504, 2051}

\bibitem[\protect\citeauthoryear{{Vink}, {de Koter}  \& {Lamers}}{{Vink}
  et~al.}{1999}]{Vink1999}
{Vink} J.~S.,  {de Koter} A.,   {Lamers} H.~J.~G.~L.~M.,  1999, \aap, \href
  {https://ui.adsabs.harvard.edu/abs/1999A&A...350..181V} {350, 181}

\bibitem[\protect\citeauthoryear{{Vink}, {de Koter}  \& {Lamers}}{{Vink}
  et~al.}{2000}]{Vink2000}
{Vink} J.~S.,  {de Koter} A.,   {Lamers} H.~J.~G.~L.~M.,  2000, \aap, \href
  {https://ui.adsabs.harvard.edu/abs/2000A&A...362..295V} {362, 295}

\bibitem[\protect\citeauthoryear{{Vink}, {de Koter}  \& {Lamers}}{{Vink}
  et~al.}{2001}]{Vink2001}
{Vink} J.~S.,  {de Koter} A.,   {Lamers} H.~J.~G.~L.~M.,  2001, \mn@doi [\aap]
  {10.1051/0004-6361:20010127}, \href
  {https://ui.adsabs.harvard.edu/abs/2001A&A...369..574V} {369, 574}

\bibitem[\protect\citeauthoryear{{Wilson}}{{Wilson}}{1915}]{Wilson1915}
{Wilson} R.~E.,  1915, \mn@doi [Lick Observatory Bulletin]
  {10.5479/ADS/bib/1915LicOB.8.132W}, \href
  {https://ui.adsabs.harvard.edu/abs/1915LicOB...8..132W} {8, 132}

\bibitem[\protect\citeauthoryear{{Yoon}, {Woosley}  \& {Langer}}{{Yoon}
  et~al.}{2010}]{Yoon2010}
{Yoon} S.~C.,  {Woosley} S.~E.,   {Langer} N.,  2010, \mn@doi [\apj]
  {10.1088/0004-637X/725/1/940}, \href
  {https://ui.adsabs.harvard.edu/abs/2010ApJ...725..940Y} {725, 940}

\bibitem[\protect\citeauthoryear{{Zucker} \& {Mazeh}}{{Zucker} \&
  {Mazeh}}{1994}]{Zucker1994}
{Zucker} S.,  {Mazeh} T.,  1994, \mn@doi [\apj] {10.1086/173605}, \href
  {https://ui.adsabs.harvard.edu/abs/1994ApJ...420..806Z} {420, 806}

\makeatother
\end{thebibliography}

% ==========================================================
\appendix
% ==========================================================

\section{Single helium-star models}
\label{sec:appendixa}

\begin{figure*}
    \includegraphics[width=\textwidth]{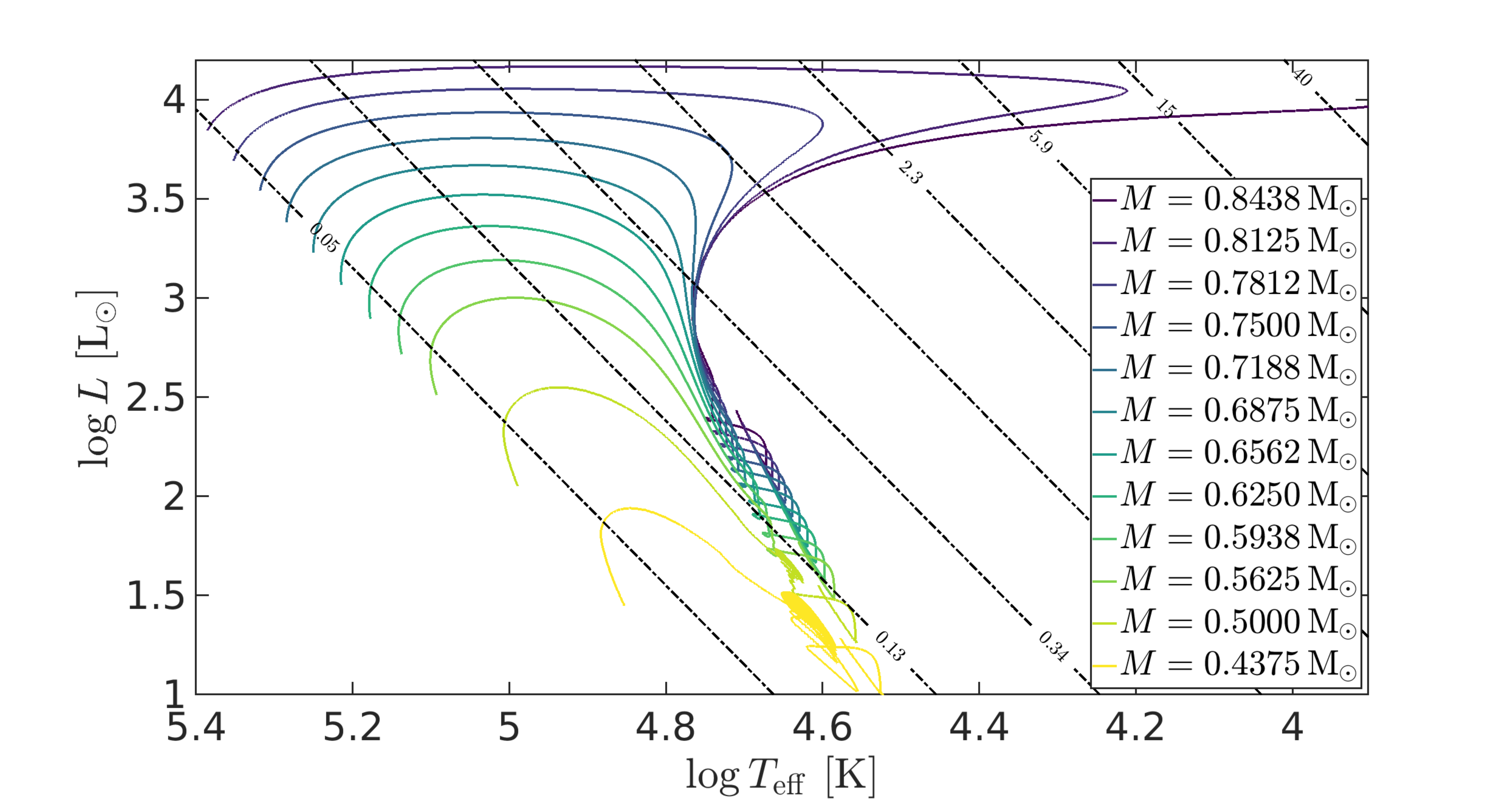}
  \caption{Hertzsprung-Russell diagram for single helium-star models.}
  \label{fig:HeliumHRD}
\end{figure*}

\begin{figure*}
    \includegraphics[width=\textwidth]{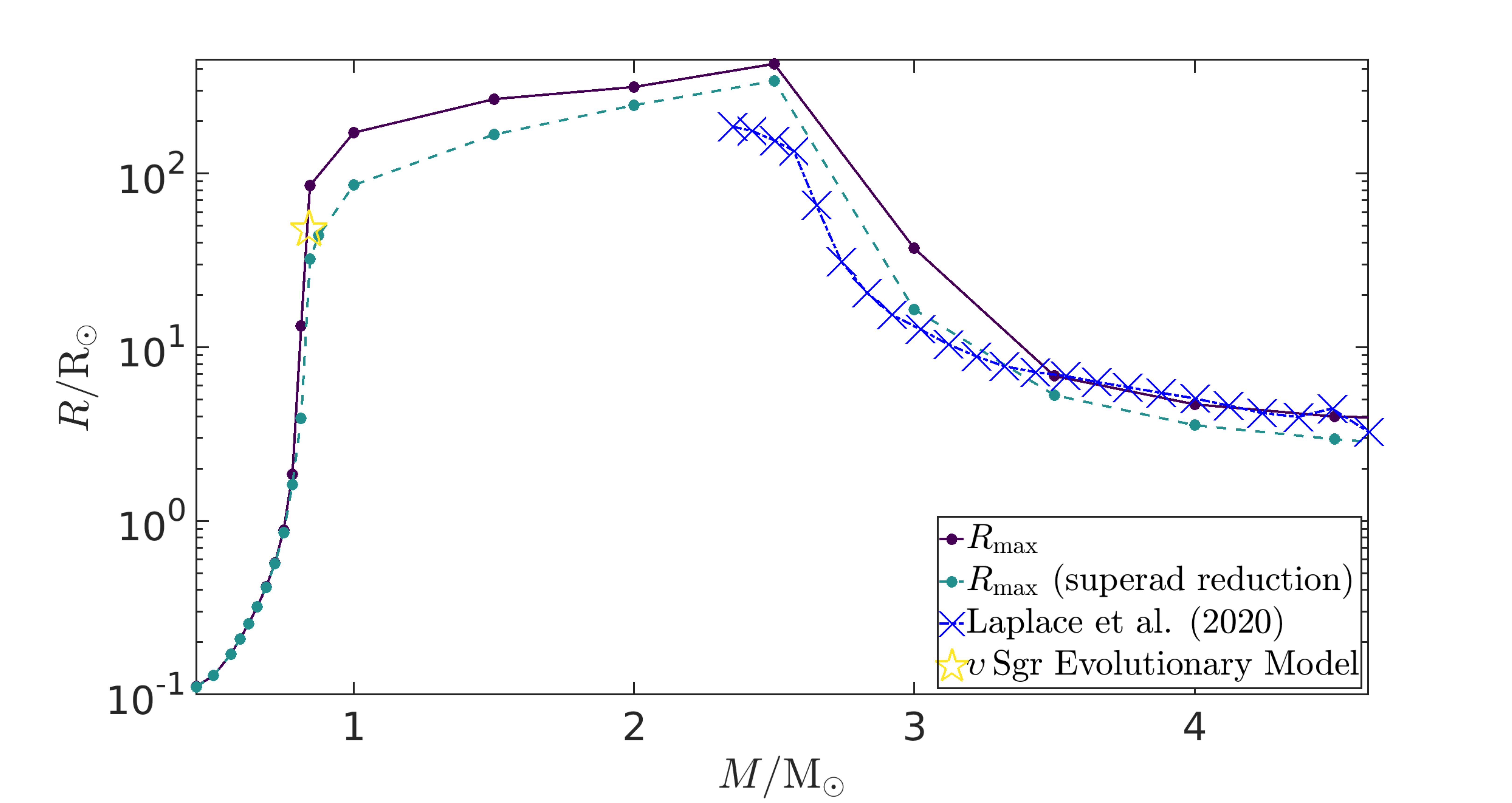}
  \caption{Maximum radial expansion in three sets of stellar evolution models compared to the best-fitting evolutionary model for $\upsilon$\,Sgr.}
  \label{fig:HeliumRmax}
\end{figure*}

\begin{figure*}
    \includegraphics[width=\textwidth]{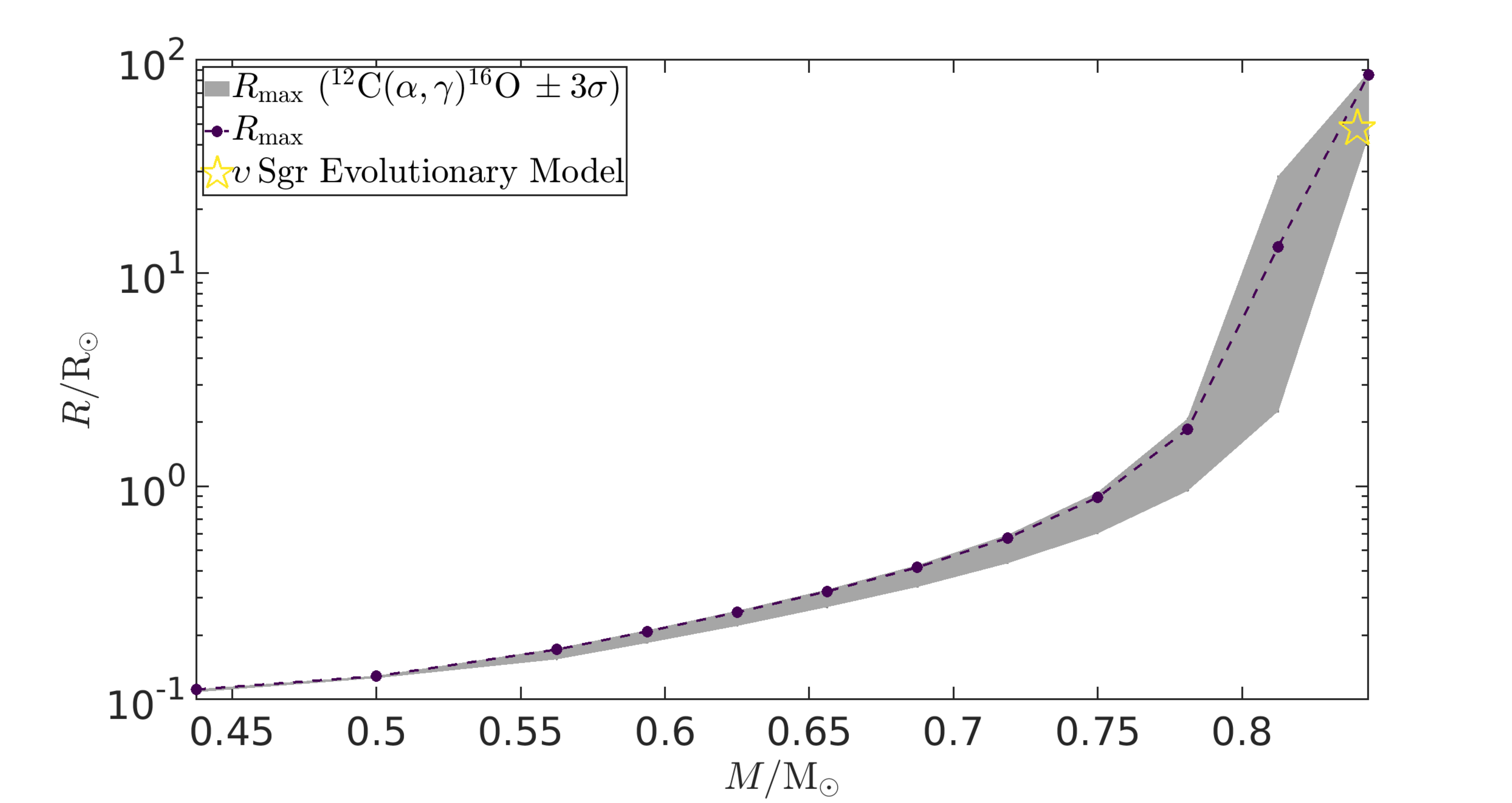}
  \caption{Maximum radial expansion in our nominal helium single-star evolution models and in models with modified $^{12}\mathrm{C}\left(\alpha , \gamma\right) ^{16} \mathrm{O}$ reaction rates compared to the best-fitting evolutionary model for $\upsilon$\,Sgr.}
  \label{fig:HeliumRmaxC12ag}
\end{figure*}

Single helium-star models are evolved by imposing mixing in the entire stellar model until the hydrogen mass fraction drops to the specified value of $X_\mathrm{H}=0.001$, with further evolution continuing with normal mixing assumptions. No wind mass loss is applied, and the surface hydrogen mass fraction stays set at $X_\mathrm{H}$ until the end of the evolution, which is either reaching $\log g = 7$, reaching a stopping condition of a central temperature of $10^{9.1}\,\mathrm{K}$, reaching the maximum model number ($15000$) or encountering numerical problems (mostly for the more massive models). Fig.\,\ref{fig:HeliumHRD} shows the evolution from the point that helium burning dominates the energy production. Fig\,\ref{fig:HeliumRmax} shows the maximum radius in the models compared to the best-fitting evolutionary model. Another set of models was evolved with the \textsc{mesa} parameter \texttt{use\_superad\_reduction} set to \texttt{.true.} (by default it is \texttt{.false.}), with the effect of suppressing density inversions resulting in smaller radii for helium-star masses of $M_\mathrm{He} \ga 0.75\,\mathrm{M}_\odot$. Our single helium-star models in the massive-star regime show good agreement with the detailed models computed by \cite{Laplace2020}. In Fig.\,\ref{fig:HeliumRmaxC12ag} we show the effect of changing the $^{12}\mathrm{C}\left(\alpha , \gamma\right) ^{16} \mathrm{O}$ reaction rate to within $\pm 3 \sigma$ of experimental data \citep{deBoer2017,Mehta2022}.
%FFFFFFFFFFFFFFFFFFFFFFFFFFFFFFFFFFFFFFFFFFFFFFFFFFFFFFFFFFFFFF
%FFFFFFFFFFFFFFFFFFFFFFFFFFFFFFFFFFFFFFFFFFFFFFFFFFFFFFFFFFFFFF
%FFFFFFFFFFFFFFFFFFFFFFFFFFFFFFFFFFFFFFFFFFFFFFFFFFFFFFFFFFFFFF
%FFFFFFFFFFFFFFFFFFFFFFFFFFFFFFFFFFFFFFFFFFFFFFFFFFFFFFFFFFFFFF
%FFFFFFFFFFFFFFFFFFFFFFFFFFFFFFFFFFFFFFFFFFFFFFFFFFFFFFFFFFFFFF
%FFFFFFFFFFFFFFFFFFFFFFFFFFFFFFFFFFFFFFFFFFFFFFFFFFFFFFFFFFFFFF
\label{lastpage}

\end{document}